\documentclass{emulateapj}

\usepackage{lscape}

\def\days{\rm day}

\shorttitle{Fraction of stars with planets in NGC 1245}
\shortauthors{Burke et~al.}

\begin{document}

\title{Survey for Transiting Extrasolar Planets in Stellar Systems:\\ 
III. A Limit on the Fraction of Stars with Planets in the Open Cluster NGC 1245}
\author{Christopher J. Burke\altaffilmark{1}, B.\ Scott Gaudi\altaffilmark{2,3}, D. L. DePoy\altaffilmark{1}, and Richard W. Pogge\altaffilmark{1}}

\altaffiltext{1}{Astronomy Department, The Ohio State University, 140 W. 18th Ave., Columbus, OH 43210}
\altaffiltext{2}{Donald H. Menzel Postdoctoral Fellow}
\altaffiltext{3}{Harvard-Smithsonian Center for Astrophysics, 60 Garden Street,
Cambridge, MA 02138}

\email{cjburke@astronomy.ohio-state.edu}

\begin{abstract} 

We analyze a 19-night photometric search for transiting extrasolar
planets in the open cluster NGC 1245.  An automated transit search
algorithm with quantitative selection criteria finds six transit
candidates; none are bona fide planetary transits. We characterize the survey
detection probability via Monte Carlo injection and recovery of
realistic limb-darkened transits. We use this
to derive upper limits on the fraction of cluster members with close-in
Jupiter-radii, $R_{J}$, companions.   The survey sample contains
$\sim$870 cluster members, and we calculate 95\% confidence upper
limits on the fraction of these stars with planets by assuming the planets
have an even logarithmic distribution in semimajor axis over the Hot
Jupiter (HJ - 3.0$<P/\days<$9.0), Very Hot Jupiter (VHJ - 1.0$<P/\days<$3.0), 
and an as of yet undetected Extremely Hot Jupiter (EHJ - $P_{\rm
Roche}<P<$1.0 day) period ranges.  For 1.5 $R_{J}$ companions we limit
the fraction of cluster members with companions to $<$1.5\%, $<$6.4\%,
and $<$52\% for EHJ, VHJ, and HJ companions, respectively.  For 1.0
$R_{J}$ companions, we find $<$2.3\% and $<$15\% have EHJ and VHJ
companions, respectively.  We do not reach the sensitivity to place
any meaningful constraints on 1.0 $R_{J}$ HJ.  From a careful analysis
of the random and systematic errors of the calculation, we show
that the derived
upper limits contain a $\pm^{13\%}_{7\%}$ relative error.  For similar
photometric noise and weather properties as this survey, observing
NGC 1245 twice as long results in a tighter constraint on HJ
companions than observing an additional cluster of similar richness as
NGC 1245 for the same length of time as this survey.  If 1\% of stars
have 1.5 $R_{J}$ HJ companions, we expect to detect one planet for every
5000 dwarf stars observed for a month.  To reach a $\sim$ 2\% upper
limit on the fraction of stars with 1.5 $R_{J}$ companions in the
3.0$<P/\days<$9.0 range, we conclude a total sample size of $\sim 7400$
dwarf stars observed for at least a month will be needed.  Results for
1.0 $R_{J}$ companions, without substantial improvement in the
photometric precision, will require a small factor larger sample size.

\end{abstract}


\section{Introduction}

Extrasolar planet detections and analysis of the non-detections
further our knowledge of the planet formation process and contribute
to an empirical determination of the typical planetary system.  These
empirical constraints will eventually decide the ubiquity or rarity of
planetary bodies in the Universe.  A variety of techniques exist to
detect extrasolar planets \citep{PER00}, and there are currently 168
extrasolar planet
discoveries\footnote{http://www.obspm.fr/encycl/catalog.html}.

Most of the extrasolar planets have been discovered using the radial
velocity technique.  Radial velocity detections indicate that
$1.2\%\pm 0.3\%$ of FGK main-sequence stars in the solar neighborhood
have a ``Hot'' Jupiter-mass planet (HJ) orbiting within 0.1 AU
\citep{MAR05}.  At this small separation from the central star, the
high temperatures and low disk column densities prevent in situ
formation of HJ planets \citep{BOD00}.  Several mechanisms exist to
exchange angular momentum between the protoplanet and natal disk,
enabling the protoplanet to migrate from a more likely formation
separation (several AU) to within 0.1 AU \citep{TER00}.  Due to tidal
circularization, HJ have nearly circular orbits with an observed
median eccentricity, $<e>=0.07$, whereas planets with larger
separations have a median eccentricity of $<e>=0.25$.  

In addition to the detection statistics and planet properties, the
extrasolar planet detections indicate several physical relationships
between the stellar host properties and the frequency of extrasolar
planets.  The most striking of these is that the probability for hosting
an extrasolar planet increases rapidly with stellar metal abundance,
consistent with $P\propto N_{\rm Fe}^{2}$ \citep{FIS05}.  The
frequency of planets may also depend on the stellar mass.
\citet{BUT04} and \citet{BON05} point out that there exists a deficit
of $M_{J}$ planets orbiting M dwarf stars.  However, the increasing
number of short-period Neptune-mass planets being discovered around M
dwarfs suggests that the overall frequency of planets (of all masses)
orbiting M dwarfs may be similar to FGK dwarfs, but the typical planet
mass is less, thereby escaping detection given the detection
limitations of the current radial velocity surveys \citep{BON05}.
Additionally, none of the M dwarfs harboring planets are metal rich
\citet{BON05}.

A coherent theory of planet formation and survival requires not only
reproducing the physical properties of the planets, but reproducing
any trends in the physical properties on the host environment.
Despite the knowledge and constraints on extrasolar planets that
radial velocity surveys provide, radial velocity surveys have their
limitations.  The high resolution spectroscopic requirements of the
radial velocity technique limit its use to the solar neighborhood and
orbital periods equivalent to the lifetime of the survey.  A full
consensus of the planetary formation process requires relying on
additional techniques to detect extrasolar planets in a larger variety
of conditions prevalent in the Universe.

For instance, microlensing surveys are sensitive to extrasolar planets
orbiting stars in the Galactic disk and bulge with distances of many
kpc away (\citealt{MAO91,GOU92}).  Two objects consistent with
Jupiter-mass companions have been detected via the microlensing
technique \citep{BON04,UDA05}.  Additional information is obtained
from studying the microlensing events that did not result in
extrasolar planet detections.  Microlensing surveys limit the fraction
of M dwarfs in the Galactic bulge with Jupiter-mass companions
orbiting between 1.5 AU to 4 AU to $<33\%$ (\citealt{ALB01,GAU02}).

Although limited to the solar neighborhood, attempts to directly image
extrasolar planets are sensitive to planets with semimajor axis beyond
20 AU.  The light from the parent star limits detecting planets
interior to the seeing disk.  Adaptive optics observations of young
($\sim 1$ Myr) stars provide the best opportunity to directly image
extrasolar planets since the young planets are still relatively bright
while undergoing a rapid, cooling contraction.  Although the
interpretation relies on theoretical modeling of these complex
planetary objects, three sources in nearby star forming regions
have been detected whose broad-band colors and spectra are consistent
with those expected from 1-42 Jupiter-mass objects \citep{NEU05,CHA05A,CHA05B}.  The contrast
ratios necessary for extrasolar planet detection are difficult to reach, and
results for detecting higher mass brown dwarfs are more complete.  An
analysis of the Cornell High-Order Adaptive Optics Survey (CHAOS)
derives a brown dwarf companion upper limit of 10\% orbiting between
25 and 100 AU of the parent star \citep{CAR05}.  \citet{MCC04}
estimate $1\%\pm 1\%$ of G,K, and M stars have brown dwarf companions
orbiting between 75 and 300 AU, but this estimate may not account for
the full range of orbital inclination and eccentricities possible
\citep{CAR05}.  At greater separations, $>$ 1000 AU, brown dwarf companions
to F-M0 main-sequence stars appear to be as common as stellar companions \citep{GIZ01}.

After the radial velocity technique, the transit technique has had the
most success in detecting extrasolar planets \citep{KON05}.  The
transit technique can detect $R_{J}$ transits in any stellar
environment where $\lesssim$1\% photometry is possible.  Thus, it
provides the possibility of detecting extrasolar planets in the full
range of stellar conditions present in the Galaxy: the Solar neighborhood,
the thin and thick disk, open clusters, the halo, the bulge, and globular clusters
are all potential targets for transit surveys.  A major advantage of
the transit technique is the current large-format mosaic CCD imagers
which provide multiplexed photometric measurements with sufficient accuracy
across the entire field of view.

The first extrasolar planet detections 
via the transit technique began with the candidate list provided by
the OGLE collaboration \citep{UDA02}.  However, confirmation of the
transiting extrasolar planet candidates requires radial velocity
observations.  Due to the well known equation-of-state competition
between electron degeneracy and ionic Coulomb pressure, the radius of
an object becomes insensitive to mass across the entire range from
below $M_{J}$ to the hydrogen-burning limit \citep{CHA00}.  Thus,
objects revealing a $R_{J}$ companion via transits may actually have a
brown-dwarf mass companion when followed up with radial velocities.
This degeneracy is best illustrated by the planet-sized brown dwarf
companion to OGLE-TR-122 \citep{PON05}.  The first radial-velocity
confirmations of planets discovered by transits \citep{KON03,BOU04}
provided a first glimpse at a population of massive, very close-in
planets with $P<$ 3 days and $M_{p}>M_{J}$ (``Very Hot Jupiters'' -
VHJ) that had not been seen by radial velocity surveys.
\citet{GAU05A} demonstrated that, after accounting for the strong
sensitivity of the transit surveys to the period of the planets, the
transit detections were likely consistent with the results from the
radial velocity surveys, implying that VHJs were intrinsically very
rare.  Subsequently, in a metallicity-biased radial velocity survey,
\citet{BOU05B} discovered a VHJ with $P=2.2$ day around the bright star
HD189733 that also has observable transits.

Despite the dependence of transit detections on radial velocity
confirmation, radial velocity detections alone only result in a lower
limit on the planetary mass, and thus do not give a complete picture
of planet formation.  The mass, radius information directly constrains the
theoretical models, whereas either parameter alone does little to
further constrain the important physical processes that shape the
planet properties \citep{GUI05}.  For
example, the mass-radius relation for extrasolar planets 
can constrain the size of the rocky core present (e.g., \citealt{LAU05}).  Also, the planet
transiting across the face of its parent star provides the exciting
potential to probe the planetary atmospheric absorption lines
against the stellar spectral features \citep{CHA02,DEM05,NAR05}.  Or,
in the opposite case, emission from the planetary atmosphere
can be detected when the planet orbits behind the parent star 
\citep{CHA05,DEM05B}.

Despite these exciting results, the transit technique is significantly hindered by the
restricted geometrical alignment necessary for a transit to occur.  
As a result, a transit survey necessarily contains at least an order of magnitude more
non-detections than detections.  In addition, null results themselves can provide
important constraints.  For example, the null result
in the globular cluster 47 Tucanae adds important empirical
constraints to the trend of increasing probability of having a
planetary companion with increasing metallicity \citep{GIL00,SAN04}.  
Thus, understanding the sensitivity of a given transit survey, i.e.\ the 
expected rate of detections and non-detections, takes on
increased importance. Several studies have taken steps toward
sophisticated Monte Carlo calculations to quantify detection
probabilities in a transit survey \citep{GIL00,WEL05,MOC05,HID05,HOO05}. 
Unfortunately, these studies do not fully characterize the sources of
error and systematics present in their analysis, and therefore the
reliability of their conclusions is unknown.  Furthermore, essentially
all of the previous studies have either (1) not accurately determined the number
of dwarf main-sequence stars in their sample, or (2) made simplifying
assumptions which may lead to misestimated detection probabilities, or (3) 
contained serious conceptual errors in the procedure with
which they have determined detection probabilities, or (4) some
combination of the above.

As a specific and important example, most studies do not apply
identical selection criteria when searching for transits amongst the
observed light curves and when recovering injected transits as part of
determining the survey sensitivity.  Removal of false-positive transit
candidates arising from systematic errors in the light curve has typically
involved subjective visual inspections, and these subjective criteria
have not been applied to the recovery of injected transits when
determining the survey sensitivity.  This is statistically incorrect,
and can in principle lead to overestimating the survey sensitivity.
Even if identical selection criteria are applied to the original
transit search and in determining the survey sensitivity, some surveys
do not apply conservative enough selections to fully eliminate
false-positive transit detections.  

In this paper, we address these shortcomings of previous studies
in our analysis of a 19-night photometric search for transiting
extrasolar planets in the open cluster NGC 1245.  An automated
transit search algorithm with quantitative selection criteria finds
six transit candidates; none are bona fide planetary transits.  
We describe our Monte Carlo calculation to robustly determine
the sensitivity of our survey, and use this to derive upper limits
on the fraction of cluster members with close-in, Jupiter-radii,
$R_{J}$, companions.

Leading up to the process of calculating the upper limit, we develop
several new analysis techniques.  First, we develop a differential
photometry method that automatically selects comparison stars to
reduce the systematic errors that can mimic a transit signal.  In
addition, we formulate quantitative transit selection criteria, which
completely eliminate false positives due to systematic light-curve variability
without human intervention.  We characterize the survey detection
probability via Monte Carlo injection and boxcar recovery of transits.
Distributing the Monte Carlo calculation to multiple processors
enables rapid calculation of the transit detection probability for a
large number of stars.

The techniques developed here enable combining results from transit
surveys in a statistically meaningful way.  This work is part of the
Survey for Transiting Extrasolar Planets in Stellar Systems (STEPSS).
The project concentrates on stellar clusters since they provide a
large sample of stars of homogeneous metallicity, age, and distance
\citep{BUR03,BUR04}.  Overall, the project's goal is to assess the
frequency of close-in extrasolar planets around main-sequence stars in
several open clusters.  By concentrating on main-sequence stars in
open clusters of known (and varied) age, metallicity, and stellar
density, we will gain insight into how these various properties affect
planet formation, migration, and survival.

The survey characteristics and the photometric procedure are given in
\S\ref{OBS}.  We explain the automated algorithm to calculate the
differential light curves and describe the light curve noise
properties in \S\ref{LC}.  In \S\ref{trandetect} we describe our
implementation of the box-fitting least squares (BLS) method
\citep{KOV02} for transit detection.  In \S\ref{sec:selcrit} we
present a thorough discussion of the quantitative selection criteria
for transit detection, followed by a discussion of the objects with
sources of astrophysical variability that meet the selection criteria
in \S\ref{trncands}.  We outline the Monte Carlo calculation for
determining the detection probability of the survey in
\S\ref{effcalc}.  We present upper limits for a variety of companion
radii and orbital periods in \S\ref{results}.  A discussion of the
random and systematic errors present in the technique is given in
\S\ref{uplimiterrsec}.  We compare the final results of this study to
our expected detection rate before the survey began and discuss the
observations necessary to reach sensitivities similar to 
radial velocity detection rates in \S\ref{discussion}.
Finally, \S\ref{conclusion} briefly summarizes this work.

\section{Observations and Data Reduction\label{OBS}}

\subsection{Observations}

We observed NGC 1245 for 19 nights between 24 Oct. and 11 Nov. of 2001
using the MDM 8K mosaic imager on the MDM 2.4m Hiltner telescope.  The
MDM 8K imager consists of a 4x2 array of thinned, 2048x4096, SITe
ST002A CCDs \citep{CRO01}.  This instrumental setup yields a
26$\arcmin$x26$\arcmin$ field of view and 0.36$\arcsec$ per pixel
resolution in 2x2 pixel binning mode.  Table~\ref{obsdat24} has an
entry for each night of observations that shows the number of
exposures obtained in the cousins $I$-band filter, median
full-width-at-half-maximum (FWHM) seeing in arcseconds, and a brief
comment on the observing conditions.  In total, 936 images produced
usable photometry with a typical exposure time of 300 s.

\subsection{Data Reduction}

We use the IRAF\footnote{IRAF is distributed by the National Optical
Astronomy Observatories, which are operated by the Association of
Universities for Research in Astronomy, Inc., under cooperative
agreement with the National Science Foundation.} CCDPROC task for all
CCD processing.  The read noise measured in zero-second images taken
consecutively is consistent with read noise measured in zero-second
images spread through the entire observing run.  Thus, the stability
of the zero-second image over the course of the 19 nights allows
median combining 95 images to determine a master, zero-second
calibration image.  For master flat fields, we median combine 66
twilight sky flats taken throughout the observing run.  We quantify
the errors in the master flat field by examining the night-to-night
variability between individual flat fields.  The small-scale,
pixel-to-pixel variations in the master flat fields are $\sim 1\%$,
and the large-scale, illumination-pattern variations reach the 3\%
level.  The large illumination-pattern error results from a
sensitivity in the illumination pattern to telescope focus.  However,
such large-scale variations do not affect differential photometry with
proper reference-star selection (as described in \S\ref{LC}).

To obtain raw instrumental photometric measurements, we employ an
automated reduction pipeline that uses the DoPHOT PSF fitting package
\citep{SCH93}.  Comparable quality light curves resulted from photometry via
the DAOPHOT/ALLFRAME, PSF-fitting photometry packages
\citep{STE87,STE98} in the background limited regime.  DoPhot performs
slightly better in terms of rms scatter in the differential light
curve in the source-noise limited regime.  The photometric pipeline
originated from a need to produce real-time photometry of microlensing
events in order to search for anomalies indicating the presence of an
extrasolar planet around the lens \citep{ALB98}.  This study uses a
variant of the original pipeline developed at The Ohio State
University and currently in use by the Microlensing Follow Up Network
\citep{YOO04}.

In brief, the pipeline takes as input a high signal-to-noise (S/N)
``template'' image.  A first pass through DoPHOT identifies the
brightest, non-saturated stars on all the images.  Using these
bright-star lists, an automated routine (J.~Menzies, private
communication) determines the geometric transformation between the
template image and all the other other images.  A second, deeper pass
with DoPhot on the template image identifies all the stars on the
template image for photometric measurement.  The photometric procedure
consists of transforming the deep-pass star list from the template
image to each frame.  These transformed positions do not vary during
the photometric solution.  Next, an automated routine (J.~Menzies,
private communication) determines an approximate value for the FWHM
and sky as required by DoPHOT.  Finally, DoPHOT iteratively determines
a best-fit, 7-parameter analytic PSF and uses this best-fit PSF to
determine whether an object is consistent with a single star, double
star, galaxy, or artifact in addition to the photometric measurement
of the object.

\section{Differential Photometry\label{LC}}

In its simplest form, differential photometry involves the use of a
single comparison star in order to remove the time variable
atmospheric extinction signal from the raw photometric measurements
\citep{KJE92}.  The process of selecting comparison stars typically
consists of identifying an ensemble of bright, isolated stars that
demonstrate long term stability over the course of the observations
\citep{GIL88}.  This procedure is sufficient for studying many
variable astrophysical sources where several percent accuracy is typically
adequate.  However, after applying this procedure on a subset of the
data, systematic residuals remained in the data that were similar
enough in shape, time scale, and depth to the expected signal
from a transiting companion to result in large number of highly-significant
false positive detections. 

Removing $\lesssim$0.01 mag systematic errors resembling a transit
signal requires a time consuming and iterative procedure for selecting
the comparison ensemble.  Additionally, a comparison ensemble that
successfully eliminates systematic errors in the light curve for a
particular star fails to eliminate the systematic errors in the light
curve of a different star.  Testing indicates each star has a
small number of stars or even a single star to employ as the
comparison in order to reduce the level of systematics in the light
curve.  On the other hand, Poisson errors in the comparison ensemble
improve as the size of the comparison ensemble increases.
Additionally, the volume of photometric data necessitates an automated
procedure for deciding on the ``best'' possible comparison ensemble.
Given its sensitivity to both systematic and Gaussian noise and its
efficient computation, we choose to minimize the standard deviation
around the mean light curve level as the figure of merit in
determining the ``best'' comparison ensemble.

\subsection{Differential Photometry Procedure}

We balance improving systematic and Poisson errors in the light curve
using the standard deviation as the figure of merit by the following
procedure.  The first step in determining the light curve for a star
is to generate a large set of trial light curves using single
comparison stars.  We do not limit the potential comparison stars to
the brightest or nearby stars, but calculate a light curve using all
stars on the image as a potential comparison star.  All comparison
stars have measured photometry on at least 80\% of the total number of
images.  A sorted list of the standard deviation around the mean
light-curve level identifies the stars with the best potential for
inclusion in the comparison ensemble.  Calculation of the standard
deviation of a light curve involves 3 iterations eliminating
3-standard-deviation outliers between iterations.  However, the
eliminated measurements not included in calculation of the standard
deviation remain in the final light curve.

Beginning with the comparison star that resulted in the smallest
standard deviation we continue to add in comparison stars with
increasingly larger standard deviations.  At each epoch, we median
combine the results from all the comparison stars making up the
ensemble after removing the average magnitude difference between
target and comparison.  We progressively increase the number of stars
in the comparison ensemble to a maximum of 30, calculating the
standard deviation of the light curve between each increase in the
size of the comparison ensemble.  The final light curve is determined
using the comparison ensemble size that minimizes the standard
deviation.  Less than 1\% of the stars result in the maximum of 30
comparison stars.  The median number of comparison stars is 4, with a
modal value of 1.  The distribution of comparison stars has a standard
deviation around the median of 4.  The fact that the standard deviations
of the majority of stars is minimized using a single comparison
star emphasizes the importance of considering all stars
as possible comparisons in order to minimize systematic errors and achieve the
highest possible accuracy.

\subsection{Comparison to a Similar Algorithm}

Independent of this study, \citet{KOV05} developed a generalized
algorithm for eliminating systematic errors in light curves 
that shares several basic properties 
with the method we have just presented.  They agree with the conclusion that
optimal selection of comparison stars can eliminate systematics in the
light curve.  They also use the standard deviation of the light curve
as their figure of merit (see their Equation 2).  However, their more
general implementation allows for the comparison star to have a real-valued,
linear correlation coefficient ($c_{i}$ in their Equation 1) in the
differential photometry, whereas the implementation outlined here forces binary
values, 0 or 1, for the linear correlation coefficient.  They solve
for the real-valued, linear correlation coefficients by minimization
of the standard deviation via matrix algebra, whereas the method given here
relies on brute force minimization of the standard deviation.

A thorough comparison of the performance between these methods has not
been done.  However, we emphasize that our algorithm found that the modal number of stars in the
comparison ensemble is a single star.  Their algorithm restricts the
comparison ensemble to a subset of the available stars.  The
restricted comparison ensemble may not capture the systematics present
in a light curve.  However, their real-valued, linear correlation
coefficients may provide the degree of freedom lacking in the
algorithm of this study necessary to cancel the systematics.  Both
algorithms possess an important caveat.  The figure of merit cannot
distinguish between improvements in the Poisson error or systematic
error and therefore does not guarantee optimal elimination of the
systematic deviations.  

\subsection{Additional Light-curve Corrections}

Although our procedure for optimally choosing comparison stars
succeeds in dramatically reducing systematics in the light
curves, we find that some additional systematic effects nevertheless
remain.  We introduce several additional corrections to the light
curves to attempt to further reduce these effects.

In good seeing, brighter stars display saturation effects.  Whereas,
in the worst seeing, some stars display light-curve deviations that
correlate with the seeing.  To correct for these effects, we fit a
two-piece, third-order polynomial to the correlation of magnitude
versus seeing.  The median seeing separates the two pieces of the fit.
We first fit the good-seeing piece with the values of the polynomial
coefficients unconstrained.  We then fit the poor-seeing piece, but
constrain the constant term such that the fit is continuous at the
median seeing.  However, we do not constrain the first or higher order
derivatives to be continuous.  In performing this fit, we excise
measurements from the light curve that would lead to a
seeing-correlation correction larger than the standard deviation of
the light curve.  We use this two-piece fit to correct the
measurements.

Measurements nearby bad columns on the detector also
display systematic errors that are not removed by the differential
photometry algorithm.  Thus, measurements when the stellar center is
within 6 pixels of a bad column on the detector are eliminated from
the light curve.

The final correction of the data consists of discarding measurements
that deviate by more than 0.5 mag from the average light curve level.  This
prevents detection of companions with radii $>3.5 R_{J}$ around the
lowest mass stars of the sample.

\subsection{Light-curve Noise Properties\label{sec:noise}}

\begin{figure}
\epsscale{1.2}
\plotone{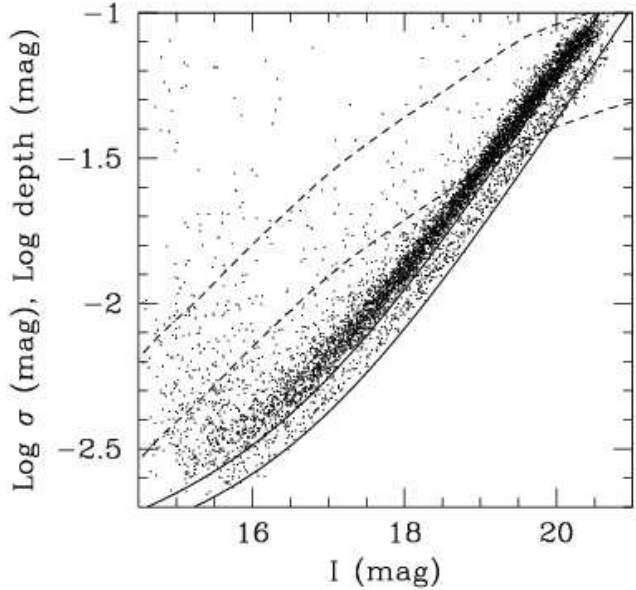}

\caption{Shows the logarithm of the light-curve standard
deviation as a function of the apparent $I$-band magnitude ({\it points}).  The
depths of transits due to a 1.0 and 1.5 $R_{J}$ companion assuming
the star is a cluster member are shown as {\it dashed lines}.  The
{\it solid lines} show photometric noise models that match the
empirically determined noise properties.\label{magrms}}

\end{figure}

Figure~\ref{magrms} shows the logarithm of the standard deviation of
the light curves as a function of the apparent $I$-band magnitude.
Calculation of the standard deviation includes one iteration with
3-standard-deviation clipping.  To maintain consistent S/N at fixed
apparent magnitude, the transformation between instrumental magnitude
to apparent $I$-band magnitude only includes a zero-point value, since
including a color term in the transformation results in stars of
varying spectral shape and thus varying S/N in the instrumental $I$
band having the same apparent $I$-band magnitude.  Each individual CCD
in the 8K mosaic has its own zero point, and the transformation is
accurate to 0.05 mag.

One CCD has significantly better noise properties than the others as
evidenced by the second sequence of points with improved standard deviation
at fixed magnitude.  The instrument contains a previously unidentified
problem with images taken in the binning mode.  The data was taken
with 2x2 native pixels of the CCD array binned to one pixel on
readout.  During readout, the control system apparently did not record
all counts in each of the four native pixels.  However, the single CCD
with improved noise properties does not suffer from this problem whereas all
the other CCDs do.  Subsequent observations with large positional
shifts allow photometric measurements of the same set of stars on the
affected detectors and unaffected detector.  Performing these
observations in the unbinned and binning modes confirms that on the
affected detectors, 50\% of the signal went unrecorded by the data
system.  This effectively reduces the quantum efficiency by half
during the binned mode of operation for seven of the eight detectors.

The two solid lines outlining the locus of points in
Figure~\ref{magrms} provide further evidence for the reduction in
quantum efficiency. These lines represent the expected noise due to a
a source-noise limited error, a term that scales as a background-noise limited error,
and 0.0015 mag noise floor.   We determine the lower line by varying the 
area of the seeing disk and the flat noise level until the noise model visually
matches the locus of points for the detector with the lower noise
properties.   Then, the upper line results from assuming half
the quantum efficiency of the lower noise model while keeping the
noise floor the same.  The excellent agreement between the higher
noise model and the noise properties of the remaining detectors
strongly supports the conclusion that half of the native pixels are
not recorded during readout.  This readout error could introduce
significant errors in the limit of excellent seeing.  However, only
4\% of the photometric measurements have FWHM$<$2.5 binned pixels.
Thus, even in the binning mode, we maintain sufficient sampling of the
PSF to avoid issues resulting from the readout error.  

The different
noise properties between detectors does not complicate the analysis.
The transit detection method involves $\chi^{2}$ merit criteria (see
\S\ref{sec:selcrit}) that naturally handle data with varying noise
properties.  Other than reducing the overall effectiveness of the
survey, the different noise properties between the detectors does not
adversely affect the results in any way.

In addition to the empirically determined noise properties, DoPhot
returns error estimates that on average result in reduced
$\chi^{2}=0.93$ for a flat light-curve model.  The average reduced
$\chi^{2}$ for all the detector agree within $10\%$.
Scaling errors to enforce reduced $\chi^{2}=1.0$ for each detector
independently has a negligible impact on the results, thus we choose
not to do so.

The upper and lower dash lines in Figure~\ref{magrms} show the transit
depth assuming the star is a cluster member for 1.5 and 1.0 $R_{J}$
companions, respectively.  In Figure~\ref{magrms}, 3671 stars have
light curves with a standard deviation less than the signal of a
transiting $R_{J}$ companion.

\section{Transit Detection}\label{trandetect}

In the previous section, we describe a procedure for generating light
curves that reduces systematic errors that lead to false-positive
transit detections.  However, systematics nevertheless remain that result in highly
significant false-positive transit detections.  This section describes
the algorithm for detecting transits and methods for eliminating false
positives based on the detected transit properties.  There are two
types of false-positives we wish to eliminate.  The first is
false-positive transit detections that result from systematic errors
imprinted during the signal recording and measurement process.  The
second type of false-positive results from true astrophysical
variability that does not mimic a transit signal.  For example,
sinusoidal variability can result in highly significant 
detections in transit search algorithms.  We specifically design the
selection criteria to trigger on transit photometric variability that
affects a minority of the measurements and that are systematically
faint.  However, the selection criteria do not eliminate
false-positive transit signals due to true astrophysical variability
that mimic the extrasolar planet transit signal we seek (grazing
eclipsing binaries, diluted eclipsing binaries, etc.).

For detecting transits we employ the box-fitting least squares (BLS)
method of \citet{KOV02}.  Given a trial period, phase of transit, and
transit length, the BLS method provides an analytic solution for the
transit depth.  We show in the appendix the equivalence of
the BLS method to a $\chi^{2}$ minimization.  Instead of using the Signal
Residue \citep[Equation 5 in][]{KOV02} or 
Signal Detection Efficiency \citep[Equation 6 in][]{KOV02}  for quantifying
the significance of the detection, we use the resulting improvement in
$\chi^{2}$ of the solution relative to a constant flux fit, as outlined in the appendix.

This section begins with a discussion of the parameters affecting the
BLS transit detection algorithm.  We set the BLS algorithm parameters
by balancing the needs of detecting transits accurately and of
completing the search efficiently.  The next step involves developing
a set of selection criteria that automatically and robustly determines
whether the best-fit transit parameters result from bona fide
astrophysical variability that resembles a transit signal.  A set of
automated selection criteria that only pass bona fide variability is a
critical component of analyzing the null-result transit survey and has
been ignored in previous analyses.

Due to the systematic errors present in the light curve, statistical
significance of a transit with a Gaussian noise basis is not
applicable.  In addition, the statistical significance is difficult to
calculate given the large number of trial phases, periods, and
inclinations searched for transits.  Given these limitations, we
empirically determine the selection criteria on the actual light
curves.  Although it is
impossible to assign a formal false alarm probability to our selection
criteria, the exact values for the selection criteria are not
important as long as the cuts eliminate the false positives while
still maintaining the ability to detect $R_{J}$ objects, and identical
criteria are employed in the Monte Carlo detection probability
calculation.

\subsection{BLS Transit Detection Parameters}

The BLS algorithm has two parameters that determine the resolution of
the transit search.  The first parameter determines the resolution of
the trial orbital periods.  The BLS algorithm \citep[as implemented
by][]{KOV02} employs a period resolution with even frequency
intervals, $\frac{1}{P_{2}}=\frac{1}{P_{1}}-\eta$, where $P_{1}$ is the
previous trial orbital period, $P_{2}$ is the subsequent (longer)
trial orbital period, and $\eta$ determines the frequency spacing
between trial orbital periods.  During implementation of the BLS
algorithm, we adopt an even logarithmic period resolution by
fractionally increasing the period, $P_{2}=P_{1}\times(1+\eta)$.  The
original implementation by \citet{KOV02} for the orbital-period
spacing is a more appropriate procedure, since even frequency
intervals maintain constant orbital phase shifts of a measurement
between subsequent trial orbital periods.  The even logarithmic period
resolution we employ results in coarser orbital phase shifts between
subsequent trial orbital periods for the shortest periods and
increasingly finer orbital phase shifts toward longer trial orbital
periods.  Either period-sampling procedure remains valid with
sufficient resolution.  We adopt $\eta=0.0025$,
which, given the observational baseline of 19 days, provides $<$10\%
orbital phase shifts for orbital periods as short as 0.5 day.

The second parameter of the BLS algorithm determines the resolution in
orbital phase by binning the phase-folded light curve.  Binning of
the data in orbital phase drastically improves the numerical
efficiency, but not without loss in determining the correct transit
properties.  \citet{KOV02} give a thorough examination of how the
sensitivity in recovering transits varies with orbital-phase binning
resolution.  To search for transit candidates in the light curves we
adopt $N_{\rm bins}=400$ orbital-phase bins.  We verify with tests
that the above parameters accurately recover boxcar signals in the
light curves.  After injection of boxcar signals in the
light curves, we calculate the $\chi^{2}$ of the solution returned by
the BLS method with the $\chi^{2}$ of the injected model.  Tests show
that the BLS method with the above parameters return a $\chi^{2}$
within 30\%, and typically much better, of the injected model's
$\chi^{2}$. 

\subsection{Selection Criteria\label{sec:selcrit}}

We apply the BLS method following the description in the previous
section to search for transit candidates in all 6787 stars with light
curves.  A visual inspection of a light curve folded at the best-fit
transit period can generally be used to discriminate between bona fide astrophysical
variability and a false positive arising from systematic errors.  However, a
proper statistical assessment of the sensitivity of a transit search
requires that the exact same set of selection criteria that are
applied to cull false positives are also applied when assessing the
detection probability via, e.g., Monte Carlo injection and recovery of
artificial signals.  Due to the large number of artificial signals
that must be injected to calculate the detection probability properly,
using selection criteria based on visual inspection of light curves is
practically very difficult or impossible.  Therefore, quantitative,
automated detection criteria that mimic the visual criteria must be
used.

We employ four selection criteria that eliminate
all false detections while still maintaining the ability to detect
$R_{J}$ companions.  These four selection criteria constitute cuts
on (1) the improvement in $\chi^2$ of a transit model over a constant flux model,
(2) the ratio between the $\Delta \chi^{2}$ of the best-fit transit
model and the $\Delta \chi^{2}_-$ of the best-fit anti-transit model, 
(3) the fraction of $\Delta \chi^{2}$ from a single night,
and (4) the transit period.

\begin{figure}
\epsscale{1.5}
\plotone{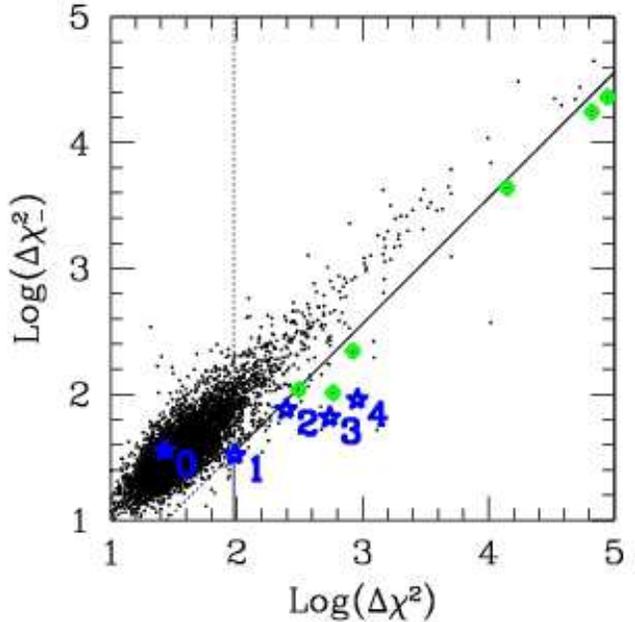}

\caption{The {\it small points} show $\Delta \chi^{2}$ as a function of $\Delta
\chi^{2}_{\rm -}$ for the resulting best-fit transit parameters in all
light curves.  Here $\Delta \chi^{2}$ and $\Delta
\chi^{2}_{\rm -}$ are the $\chi^{2}$ improvement between the flat
light-curve model and the best-fit transit and anti-transit model,
respectively.  The {\it dotted vertical line} shows the $\Delta
\chi^{2}=95.0$ selection boundary.  The {\it solid diagonal line}
shows the $\Delta \chi^{2}/\Delta \chi^{2}_{\rm -}=2.75$ selection
boundary.  Objects in the lower right corner pass both selection
criteria.  The {\it green diamonds} show values of $\Delta \chi^{2}$
and $\Delta \chi^{2}_{\rm -}$ for the six transit candidates.  The
{\it blue stars} show the recovered values of $\Delta \chi^{2}$ and
$\Delta \chi^{2}_{\rm -}$ for the four light curves with injected
transits shown in Figure~\ref{falsetran}.  The label next to the blue
stars corresponds to the label in the upper right corner of each panel
in Figure~\ref{falsetran}.  These curves were created by injecting
transits into the same light curve.  The blue star labeled 0 shows the
values of $\Delta \chi^{2}$ and $\Delta \chi^{2}_{\rm -}$ for this
light curve before the example transits were injected.\label{selcrit}}

\end{figure}

The first of the selection criteria is a cut on
$\Delta \chi^{2}$, the improvement in $\chi^{2}$ between a
constant flux model and a transit model.  The $\Delta \chi^{2}$ is
similar to the Signal Residue, SR, of \citet{KOV02}; we derive $\Delta
\chi^{2}$ and its relation to SR in the appendix.  We prefer $\Delta
\chi^{2}$ over SR as the former allows a direct comparison of the
transit detection significance between light curves with different
noise properties.  Given the correlated systematics in the data, we
cannot rely on analytical formulations with a Gaussian statistics
basis for the statistical significance of a particular $\Delta
\chi^{2}$ value.  We empirically determine a cut on $\Delta \chi^{2}$ in
combination with the other selection criteria to fully eliminate false
detections.  For a transit detection we require $\Delta
\chi^{2}>95.0$.  As shown in the appendix, this selection criterion
corresponds to a S/N$\sim$10 transit detection.  Figure~\ref{selcrit}
shows the $\Delta \chi^{2}$ of the best fit transit for all light
curves along the x-axis.  The vertical line designates the selection
criteria on this parameter.  Even with such a strict
threshold, there are still a large number of false
positives that pass the $\Delta \chi^2$ cut.

Systematic variations in the light curves that are characterized by
small reductions in the apparent flux of star that are coherent over
the typical time scales of planetary transits can give rise to
false-positive transit detections.  However, under the reasonable
expectation that systematics do not have a strong tendency to produce
dimming versus brightening of the apparent flux of the stars, one
would expect systematics to also result in a false-positive
`anti-transit' (brightening) detections.  Furthermore, most intrinsic
variables can be approximately characterized by sinusoids, which will
also result in significant transit and anti-transit detections.  On the other
hand, a light curve with a true transit signal and insignificant systematics
should produce only a strong transit detection, and not a strong anti-transit
detection.

Thus, the ratio of the significance of the best-fit transit signal
relative to that of the best-fit anti-transit signal provides a rough
estimate of the degree to which a detection has the expected
properties of a bona fide transit, rather than the properties of
systematics or sinusoidal variability.  In other words, a highly
significant transit signal should have a negligible anti-transit
signal, and therefore we require the best-fit transit to have a
greater significance than the best-fit anti-transit.  We accomplish
this by requiring transit detections to have $\Delta \chi^{2}/\Delta
\chi^{2}_{\rm -}>2.75$, where $\Delta \chi^{2}_{\rm -}$ is the
$\chi^{2}$ improvement of the best-fit anti-transit.  For a given
trial period, phase of transit, and length of transit, the BLS
algorithm returns the best-fit transit without restriction on the sign
of the transit depth.  Thus, the BLS algorithm simultaneously searches
for the best-fit transit and anti-transit, and so determining $\Delta
\chi^{2}_{\rm -}$ has no impact on the numerical efficiency.

Figure~\ref{selcrit} shows the $\Delta \chi^{2}_{\rm -}$ of
the best fit anti-transit versus the $\Delta \chi^{2}$ of the
best-fit transit for our light curves.  The diagonal line demonstrates the selection on
the ratio $\Delta \chi^{2}/\Delta \chi^{2}_{\rm -}=2.75$.  Objects
toward the lower right corner of this Figure pass the selection
criteria.  The objects with large $\Delta \chi^{2}$ typically have
correspondingly large $\Delta \chi^{2}_{\rm -}$.  This occurs for
sinusoidal-like variability or strong systematics that generally have 
both times of bright and faint measurements with respect to the mean light-curve level.

\begin{figure}
\epsscale{1.5}
\plotone{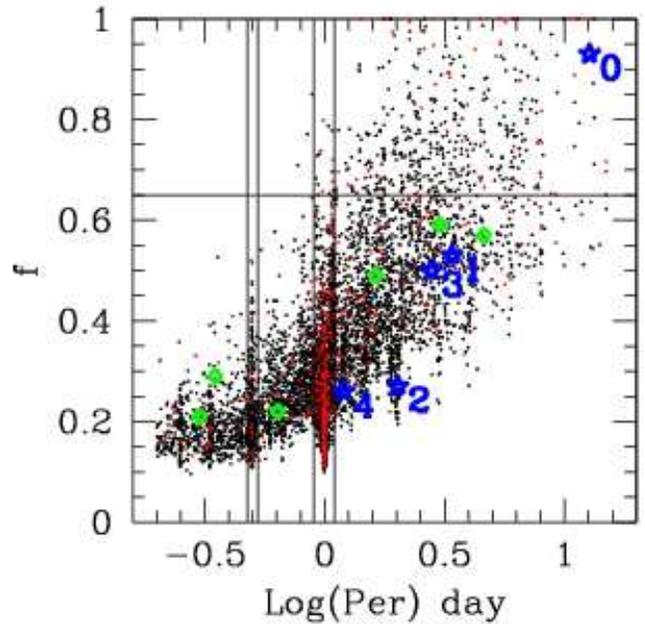}

\caption{The {\it black points} show $f$ as a function of the best-fit transit orbital period,
where $f$ is the fraction of the total $\chi^{2}$ improvement with the
best-fit transit model that comes from a single night.  The objects that pass the $\Delta \chi^{2}>95.0$ selection
criteria are shown as {\it red points}.  The {\it horizontal line}
shows the $f=0.65$ selection boundary.  The {\it vertical lines}
denote orbital period regions avoided due to false-positive transit
detections.  The {\it blue stars} and {\it green diamonds} are the
same as in Figure~\ref{selcrit}.\label{selcrit2}}

\end{figure}

Requiring observations of the transit signal on separate nights also
aids in eliminating false-positive detections.  We quantify the
fraction of a transit that occurs during each night based on the
fraction of the transit's $\chi^{2}$ significance that occurs during
each night.  The parameters of the transit allow identification of the
data points that occur during the transit.  We sum the individual
$\chi^{2}_{i}=\left(m_{i}/\sigma_{i}\right)^{2}$ values for data
points occurring during the transit to derive $\chi^{2}_{\rm tot}$,
where $m_{i}$ is the light curve measurement and $\sigma_{i}$ is its
error.  Then we calculate the same sum for each night individually.
We denote this $\chi^{2}_{k\rm th\: night}$.  We identify the night
for which $\chi^{2}_{k\rm th\: night}$ contributes the greatest
fraction of $\chi^{2}_{\rm tot}$, and we call this fraction
$f=\chi^{2}_{k\rm th\: night}/\chi^{2}_{\rm tot}$.  Finally, we
require $f <0.65$.  This corresponds to roughly seeing the transit one
and a half times assuming all observations have similar noise.
Alternatively, this criterion is also met by observing $2/3$ of a
transit on one night and $1/3$ of the transit on a separate night, or
observing a full transit on one night and $1/6$ of transit on a
separate night with three times improvement in the photometric error.
Figure~\ref{selcrit2} shows $f$ versus the best-fit period for all the
light curves.  The horizontal line designates the selection on this
parameter.

The red points in Figure~\ref{selcrit2} show objects that pass the
$\Delta \chi^{2}>95.0$ selection.  We find that most are clustered
around a 1.0 day orbital period.  A histogram of the best-fit transit
periods amongst all light curves reveals a high frequency for 1.0 day
and 0.5 day periods.  Visual inspection of the phased light curves
reveals a high propensity for systematic deviations to occur on the
Earth's rotational period and 0.5 day alias.  We do not fully
understand the origin of this effect, but we can easily conjecture on
several effects that may arise over the course of an evening as the
telescope tracks from horizon to horizon following the Earth's diurnal
motion.  In order to eliminate these false positives, we apply as our
fourth selection criteria a cut on the period.  Specifically, we
require transit detections to have periods that are not within $1.0\pm
0.1$ and $0.5\pm 0.025$ day.  The horizontal lines designate these
ranges of discarded periods.

\section{Transit Candidates}\label{trncands}

\begin{figure}
\epsscale{1.2}
\plotone{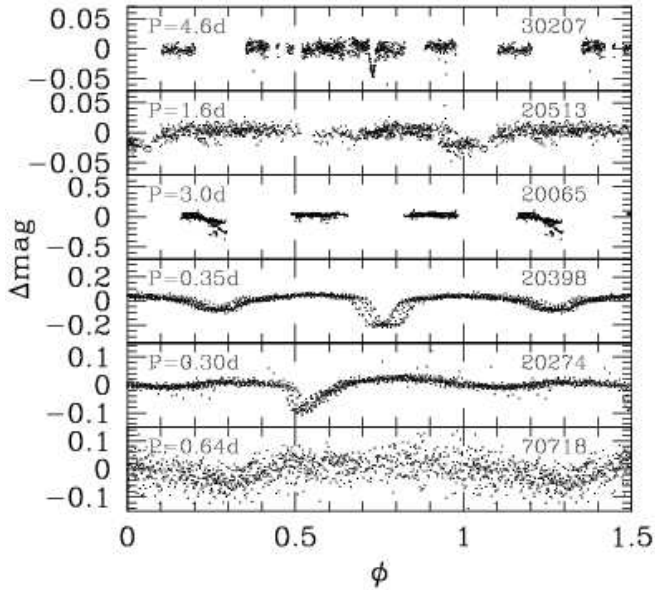}

\caption{The points show the change in magnitude as a function of orbital phase for all
stars that meet the transit candidate selection criteria.  Negative
values for $\Delta$ mag are toward fainter flux levels.  The phased
period is given in the upper left corner of each panel, and the number
in the upper right corner of each panel gives the internal
identification number.\label{trncand}}

\end{figure}

Six out of 6787 stars pass all four selection criteria.  All
of these stars are likely real astrophysical variables whose
variability resembles that of planetary transit light curves.
However, we find that none are bona fide planetary transits in NGC
1245.  After describing the properties of these objects we will
describe the procedure for ruling out their planetary nature.
Figure~\ref{trncand} shows the phased light curves for these six stars.
Each light-curve panel in Figure~\ref{trncand} has a different
magnitude scale with fainter flux levels being more negative.  The
upper left corner of each panel gives the detected transit period as
given by the BLS method.  The upper right corner of each panel gives
an internal identification number.  The panels from top to bottom have
decreasing values in the ratio between the improvement of a transit
and anti-transit model, $\Delta \chi^{2}/\Delta \chi^{2}_{\rm -}$.

\begin{figure}
\epsscale{1.2}
\plotone{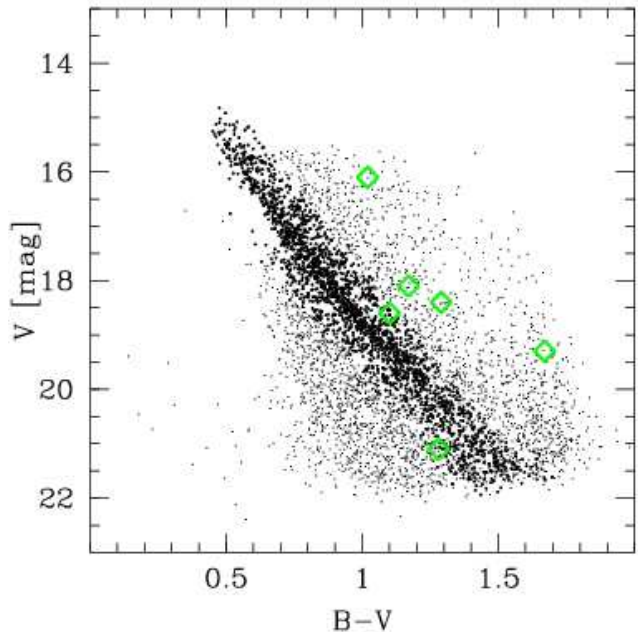}

\caption{The CMD of the cluster NGC 1245.  Potential cluster members
having $\chi^{2}_{\rm mem}<0.04$ are shown with {\it heavy points}).
Objects that exceed the selection criteria for transit detection are
given as {\it open diamonds}.\label{cmd}}

\end{figure}

Table~\ref{trncandprop} lists the properties and selection criteria
values for the stars shown in Figure~\ref{trncand}.  The green
diamonds in Figures~\ref{selcrit} and \ref{selcrit2} represent the
selection criteria for the six transit candidates.  The photometric
and positional data in Table~\ref{trncandprop} come from
\citet{BUR04}.  The $\chi^{2}_{\rm mem}$ entry in
Table~\ref{trncandprop} measures the photometric distance of a star
from the isochrone that best fits the cluster CMD.  A lower value of
this parameter means a star has a position in the CMD closer to the
main sequence.  Heavy points in Figure~\ref{cmd} denote stars with
$\chi^{2}_{\rm mem}<0.04$, and we designate these stars as potential
cluster members.  Based on $\chi^{2}_{\rm mem}$, star 20513 and star
70178 have photometry consistent with cluster membership, thus we also
list the physical parameters of those stars in
Table~\ref{trncandprop}.  \citet{BUR04} details the procedure for
determining the physical parameters of a star based solely on the
broad-band photometry and the best-fit cluster isochrone.  However,
the validity of the stellar physical parameters only applies if the
star is a bona fide cluster member.

\begin{figure}
\epsscale{0.8}
\plotone{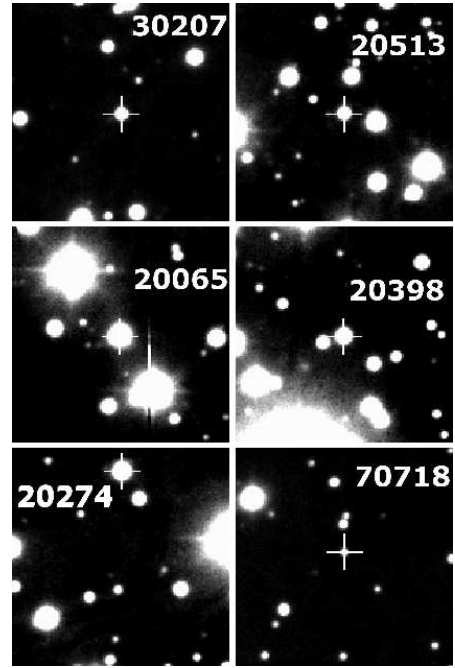}

\caption{Finding charts for transit candidates with light curves shown in
Figure~\ref{trncand}.  Each panel is 54$\arcsec$ on a side.  North is
toward the right, and East is toward the bottom.\label{find}}

\end{figure}

Figure~\ref{find} shows a finding chart for each star with a light
curve in Figure~\ref{trncand}.  The label in each panel gives the
identification number, and the cross indicates the corresponding
object.  Star 20274 is not centered in the finding chart because it is
located near the detector edge.  The field of view of each panel is
54\arcsec.  North is toward the right, and East is toward the bottom.
The panels for stars 20065, 20398, and 20513 (located near the cluster
center) provide a visual impression of the heaviest stellar crowding
encountered in the data.  Figure~\ref{cmd} shows the $V$ and $\bv$ CMD
of the cluster field as given in \citet{BUR04}.  The large open stars
denote the locations of the objects that exceed the transit selection
criteria.

\subsection{Consistency of Transit Parameters with Cluster Membership}

Only stars 20513 and 70718 have $\chi^{2}_{\rm mem}$ values consistent
with cluster membership.  Additionally, the transit depth in both
stars indicates potential for having a $R_{J}$ companion.  However,
qualitatively, in each case the transit duration relative to the
orbital period is too long to be a true planetary companion to a
cluster main-sequence star.  We can use our knowledge of the physical
properties of the parent stars to quantitatively rule out planetary
companions.  We do this by comparing an estimate of the stellar radius
derived from the CMD to an independent estimate of a lower limit on
the stellar radius derived from the properties of the light curve.  In
both cases we find that the stellar radii derived from the CMD are
well below the lower limit on the stellar radius
based on the light curve.

To derive a lower limit on the stellar radius from the light curve,
 we build on the work by \citet{SEA03}.  They provide 
a purely geometric relationship between the
orbital semimajor axis, $a$, and stellar radius, $R_{\star}$, for a light curve
with a given period, $P$, depth of transit, $\Delta F$, and total
duration of the transit (first to fourth contact), $\tau$, assuming
a circular orbit (see their Equation 8).  
By assuming a central transit (impact parameter $b=0$), we transform their equality 
into a lower limit.  Using Kepler's Third Law, assuming that 
the mass of the companion is much smaller than the mass of the star,
and assuming the duration of the transit is much smaller than the period ($\tau \ll P$),
we find,
\begin{equation}
R_{\star}>\frac{\pi (M_{\star}+m_{p})^{1/3}\tau}{P\,^{1/3}(1+\sqrt{\Delta F})}.
\label{eqn:rstar}
\end{equation}
where $R_{\star}$ is in AU, $M_{\star}$ is in units of $M_{\odot}$ and $\tau$ and
$P$ are in years. 

Parameters on the right hand side of the above equation contain
substantial uncertainties.  Replacing the parameters by their maximum
plausible deviation from their measured values in such a manner as to
decrease $R_{\star}$ increases the robustness of the lower limit.  The
orbital period determination has the largest
uncertainty.  Tests of recovering transits in the light curves reveal
a 10\% chance for the BLS method to return an orbital period, $P'$, at
the $1/2P$ and $2P$ aliases of the injected orbital period, and a
$<1$\% chance of detecting the $1/3P$ and $3P$ aliases.
Misidentification of the correct orbital period results from gaps in
the observing window function.  Replacing $P$ in the above equation
with $3P'$, where $P'$ is the orbital period returned by the BLS
algorithm, provides the maximum plausible deviation of this quantity
and increases the robustness of the lower limit.  In addition, the
stellar mass determination based on the CMD
potentially has contamination from a binary companion.  Thus, we
replace $M_{\star}$ with $0.5 M'_{\star}$, where $M'_{\star}$ is the
stellar mass estimate from the CMD.  We do not
modify $\tau$ or $\Delta F$.  For the cases considered here, $\Delta F \ll 1$,
and the term $1+\sqrt{\Delta F} \simeq 1$ in Equation \ref{eqn:rstar}.
Therefore the precise value of $\Delta F$ has little effect on
the resulting limit on $R_*$.  The BLS algorithm fits a boxcar transit
model to the light curve via a $\chi^{2}$ minimization.  Since, in the
limit of zero noise, any non-zero boxcar height fit to a transit can
only result in an increasing $\chi^{2}$ when the length of the boxcar
exceeds the length of the transit, $\tau$ underestimates the true
transit length.  Making the above replacements the lower limit on the
stellar radius is,
\begin{equation}
R_{\star}>7.3\frac{(M'_{\star}/M_{\odot})^{1/3}(\tau/\rm{1\: day})}{(P'/\rm{1\: day})^{1/3}(1+\sqrt{\Delta F})} R_{\odot}.
\end{equation}

For star 20513, the above equation requires $R_{\star}>1.04 R_{\odot}$
if the star is a cluster member.  Fits to the CMD yield a stellar
radius $R_{\star}=0.80 R_{\odot}$.  The lower limit for star 70718
is $R_{\star}>0.82 R_{\odot}$, whereas the CMD yields $R_{\star}=0.56
R_{\odot}$.  Clearly both stars lack consistency between the stellar
radius based on the CMD location and the stellar radius based on the
transit properties.

The transiting companions to  20513 and 70718 are also unlikely
to be planets if the host stars are field dwarfs.
\citet{TIN05} provide a
diagnostic to verify the planetary nature of a transit when only the
light curve is available.  The diagnostic $\eta_{p}$ of \citet{TIN05} compares
the length of the observed transit to an estimate of the transit
length derived by assuming a main-sequence mass-radius relation for the central star.  By
assuming a radius of the companion of $R_{p}=1.0 R_{J}$, we find
$\eta_{p}=4.0$ and $3.8$ for 20513 and 70718, respectively.  Values of
$\eta_{p}\lesssim 1$ correspond to planetary transits.  Therefore, 
20513 and 70718 are unlikely to host planetary companions
with $R_{p} \la R_{J}$ if they are main-sequence stars.

We note that our final {\it a posteriori} criterion with which we reject cluster
transit candidates, namely the consistency between the radius of
the parent star as estimated from the CMD and the radius as estimated
from the light curve, is a conceptually different kind of selection
criterion than those that we applied to all the light curves to arrive
at our six transit candidates.  The original four selection criteria
were designed to detect bona fide astrophysical variability that
resembles the signals from transiting planets, but does not
necessarily arise from a transiting planetary companion.  In
principle, we could have included the radius consistency cut as an
additional selection criterion applied to all light curves.  The
motivation to do this would be that imposing this additional criterion
might automatically remove some systematic false positives and so
allow us to improve our efficiency by making the other selection
criteria less stringent.  We have found using limited tests that
this is not the case.  We therefore chose to leave the radius check as
an {\it a posteriori} cut on the transit candidates.  Nevertheless,
observing a cluster does provide an advantage over observing field
stars, as the additional constraint on the stellar radius from the
cluster CMD provides a more reliable confirmation of the planetary
nature than the light curve alone \citep{TIN05}, and furthermore
allows a more accurate assessment of the detection probability.

It is important to emphasize that all of the injected transits with
which we compute the detection probability (\S\ \ref{effcalc})
automatically pass the radius consistency criterion.  A fraction of
these will be recovered at periods that differ enough from the input
period that by using the recovered period they will no longer satisfy
the radius constraint.  However, we find that this fraction is
negligibly small.

\subsection{Individual Cases}

This section briefly discusses each object that met the selection
criteria as a transit candidate but does not belong to the cluster.
The V-shaped transit detected in star 30207 rules out a $R_{J}$
companion.  Transiting $R_{J}$ companions result in a flat
bottomed eclipse as the stellar disk fully encompasses the planetary
disk.  A closer inspection of the light curve also reveals ellipsoidal
variations outside of the transit.  This light curve matches the
properties of a grazing eclipse, which is a typical contaminant in
transit searches (e.g., \citealt{BOU05}).

The remaining stars have depths too large for a $R_{J}$ companion and
show evidence for secondary eclipses.  Recall that we eliminated 
data points with $|\Delta m|> 0.5$ mag in the light curves.  This eliminates the eclipse
bottom for star 20065.  Keeping all the data for star 20065 clearly
reveals the characteristics of a detached eclipsing binary.
The period BLS derives for star 20065 aligns the primary
and secondary eclipses, and thus BLS-reported period is not the true orbital
period.

The eclipses in stars 20398 and 20274 do not perfectly phase up.  This
is because the resolution in period we used for the search prevents
perfect alignment of the eclipses for such short periods.  This effect
is inconsequential for detecting transiting planets as they all have
orbital periods longer than 0.3 day.

Finally, we note that other
variables exist in the dataset.  They were not selected
because they do not meet the $\Delta
\chi^{2}_{\rm min}/\Delta \chi^{2}_{\rm min\, -}$ selection criterion.
A future paper will present variables that exist in this dataset using
selection criteria more appropriate for identifying quasi-sinusoidal periodic
variability (J. Pepper et al., in preparation).

\section{Detection Probability Calculation}\label{effcalc} 

We did not detect any transit signals consistent with a $R_{J}$
companion.  To interpret this null result in terms of the frequency of
planetary companions to stars in NGC 1245, we develop a Monte Carlo
detection probability calculation for quantifying the sensitivity of
the survey for detecting extrasolar planet transits.  The calculation
provides the probability of detecting a transit in the survey as a
function of the companion semimajor axis and radius.  In addition to
the photometric noise and observing window, the observed properties of
the transit signal depend sensitively on the host mass, radius,
limb-darkening parameters, and orbital inclination with respect to the
line of sight.  Without accurate knowledge of the stellar parameters,
a detailed detection probability is not possible.  This precludes
analyzing stars not belonging to the cluster.  Given the degeneracy
between broad-band colors of dwarfs, subgiants, and giants, the
stellar radius for most field objects cannot be determined from the
CMD alone.  Assuming all stars of a given color are dwarfs drastically
overestimates the number of actual dwarf stars in a transit survey
\citep{GOU03}.  The minimal expenditure of observational resources
necessary for determining the stellar parameters for a cluster transit
survey provides a significant advantage over transit surveys of the
field.

Each star in the survey has a unique set of physical properties and
photometric noise, thus we calculate the detection probability for all
stars in the survey.  This is the first study of its kind to do so.
Given the detection probability for each star, the distribution of
extrasolar planet semimajor axis, and frequency of extrasolar planet
occurrence, the survey should have detected,
\begin{equation}
N_{\rm det}=f_{\star} \sum_{i=1}^{N_{\star}}P_{\rm det,i},
\label{eqn:ndet}
\end{equation}
extrasolar planets, where the sum is over all stars in the survey,
\begin{equation}
P_{\rm det,i}=\int \int \frac{d^{2}p}{dR_{p} da}P_{\epsilon,i}(a,R_{p})P_{T,i}(a,R_{p})P_{\rm mem,i}dR_{p}da\label{ndeteq},
\end{equation}
$R_{p}$ is the extrasolar planet radius, $a$ is the semimajor axis,
$f_{\star}$ is the fraction of stars with planets distributed
according to the joint probability distribution of $R_{p}$ and $a$,
$\frac{d^{2}p}{dR_{p} da}$.  The Monte Carlo detection probability
calculation provides $P_{\epsilon,i}(a,R_{p})$,the probability of
detecting a transit in a given light curve.  The term
$P_{T,i}(a,R_{p})$ gives the probability for the planet to cross the
limb of the host along the line of sight, and $P_{\rm mem,i}$ gives
the probability the star is a cluster member.  This framework for
calculating the expected detections of the survey follows from the
work of \citet{GAU02}.  In the following subsections we describe the
procedure for calculating each of these probability terms.

\subsection{Calculating $P_{\epsilon,i}(a,R_{p})$}

$P_{\epsilon,i}(a,R_{p})$ is the probability of detecting a transit
around the $i$th star of the survey averaged over the orbital phase
and orbital inclination for a given companion radius and semimajor
axis.  We begin this section with a description of the procedure for
injecting limb-darkened transits into light curves for recovery.
After injecting the transit, we attempt to recover the transit
employing the same BLS algorithm and selection criteria as employed
during the transit search on the original data.  It is critical to
employ identical selection criteria during the recovery as
the original transit search since only then can we trust the
robustness and statistical significance of the detection.  The
fraction of transits recovered for fixed semimajor axis and $R_{p}$
determines $P_{\epsilon}$.  Next, we characterize the sources of error
present in $P_{\epsilon}$ and how we ensure a specified level of
accuracy.  Finally, in this section we discuss the parallelization of
the calculation to obtain $P_{\epsilon}$ for all stars in the survey
in a reasonable amount of time.

In the appendix, we discuss the importance of injecting realistic
transits for recovery.  \citet{MAN02} provide analytic formulas for
calculating realistic limb-darkened transits.  We employ the
functional form of a transit for a quadratic limb-darkening law as
given in Section 4 of \citet{MAN02}.  The quadratic limb-darkening
coefficients come from \citet{CLA00}.  Specifically, we use the
$I$-band limb-darkening coefficients using the ATLAS calculation for
$\log g=4.5$, $\log$[M/H]=0.0, and $v_{turb}=2$ kms$^{-1}$.  

We assume circular orbits for the companions.  All known extrasolar
planets to date that orbit within 0.1 AU have eccentricities $<$0.3,
and the average eccentricity for these planets is
$<e>=0.07$\footnote{http://www.obspm.fr/encycl/catalog.html}.

After injecting the transit, we employ the BLS algorithm to recover
the injected transit signal using the selection criteria described in
\S\ref{sec:selcrit}.  For numerical efficiency, we relax the resolution of
the BLS search parameters.  We adopt a fractional period step,
$\eta=0.004$, and phase space binning, $N_{\rm bins}=300$.  Despite
the reduced resolution, higher resolution, converged solutions reveal
only a 0.003 lower probability resulting from the adopted parameters.
We correct all probabilities for this systematic even though it is at
an insignificant level compared to the other uncertainties.

\begin{figure}
\plotone{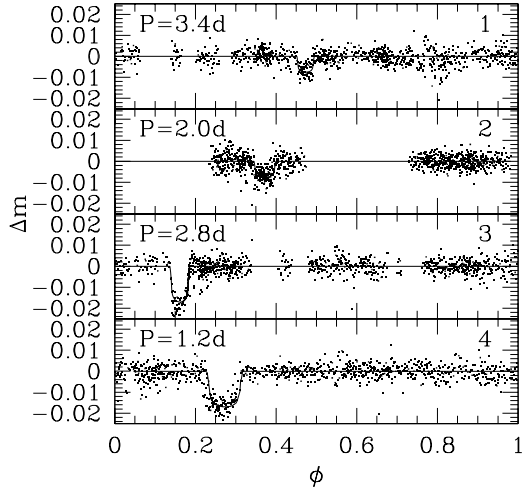}

\caption{Phased light curves showing the recovery of transits injected
in the light curve by the Monte Carlo calculation.  The injected
limb-darkened transit signal is given by the {\it solid line}.  The
top two panels and bottom two panels show results for 1.0 and 1.5
$R_{J}$ companions, respectively.  The transit recovery in the top
panel barely meets the selection criteria and gives an impression for
the sensitivity of the survey.  The labels in the upper right corner
of the panels correspond to the markers for the selection criteria
values shown in Figures~\ref{selcrit}
and~\ref{selcrit2}.\label{falsetran}}

\end{figure}

Figure~\ref{falsetran} visualizes the injected transits with
increasing degrees of significance from top to bottom.  This Figure
shows light curves with an injected transit phased at the period as
returned from the BLS algorithm.  The solid line illustrates the
injected limb-darkened transit signal.  The top two panels and the
bottom two panels illustrate 1.0 and 1.5 $R_{J}$ companions,
respectively.  The transit recovery in the top panel barely
meets the selection criteria, thus giving a visual impression for the
sensitivity of the survey.  The resulting selection criteria values
after recovery of the injected transits are shown in
Figures~\ref{selcrit} and \ref{selcrit2} by the blue stars, and the
labels next to the stars correspond to the panel label given in the
upper right hand corner. 
The modeled transits shown in this Figure are injected into the 
same light curve of a potential cluster member with $V=16.6$ and rms scatter 
(before transit injection) of $\sigma=0.003$.
The blue star, labeled 0 in Figures~\ref{selcrit}
and \ref{selcrit2}, represents the values for the selection criteria for 
found for this light curve before injecting the transits.

As opposed to previous work, we carefully examine, quantify, and control the
uncertainties present in the calculation.  During injection of a transit at
fixed semimajor axis, the transit can occur during any phase of the
orbit.  We use the following procedure
to ensure that we inject enough trial transits at random orbital 
phases to yield convergence of $P_{\epsilon}$.
Based on binomial statistics, the error in the resulting probability
at fixed orbital period depends on the actual probability and the
number of trial transit phases, $\sigma_{\epsilon}=\sqrt{N_{\rm
trial}\epsilon_{\rm act}(1-\epsilon_{\rm act})}$, where $N_{\rm
trial}$ is the number of trial transit phases and $\epsilon_{act}$ is
the actual probability (unknown a priori).  Maintaining the same error
in the detection probability for differing $\epsilon_{\rm act}$
requires a variable number of trial phases.  For each semimajor axis, 
we first obtain an initial
estimate for the probability, $\epsilon_{\rm est}$, using $N_{\rm
trial}=100$.  We then increase $N_{\rm
trial}$ until the probability converges to 
$\sigma_{\epsilon}=\sqrt{N_{\rm trial}\epsilon_{\rm
est}(1-\epsilon_{\rm est})}\leq 0.02$.  The above procedure
systematically overestimates $\epsilon_{\rm act}$ when $\epsilon_{\rm
act}\ga 0.95$ and systematically underestimates $\epsilon_{\rm act}$
when $\epsilon_{\rm act}\la 0.05$.  However, these errors
are of order the adopted $\sigma_{\epsilon}=0.02$ accuracy, and so
we neglect them here.

In addition to a random orbital phase, assuming a random orientation
of the orbit requires taking into account an even distribution in
$\cos i$, where $i$ is the inclination of the orbit.  Only a narrow
range of inclinations, $\cos i\leq (R_{\star}+R_{p})/a$, results in a
transit.  Thus, we inject the transit with an even distribution in
$\cos i$ between $0\leq \cos i \leq (R_{\star}+R_{p})/a$.

The previous discussion pertains to ensuring a prescribed accuracy at
fixed semimajor axis.  However, the expected detection rate also
requires an integral over semimajor axis, which must be sampled at
high enough resolution to ensure convergence of the integral.  We
calculate the probability at even logarithmic intervals, $\delta \log
a=0.011$ AU.  In comparison to high-resolution, converged
calculations, this semimajor axis resolution results in an absolute
error in the detection probability integrated over the semimajor axis
of $\sigma_{\epsilon}=0.003$.  We inject transits with semimajor axis
from the larger of 0.0035 AU and $1.5 R_{\star}$ to 0.83 AU.  The
best-fit isochrone to the cluster CMD determines the parent star
radius.

Generating the light curve from the raw photometric measurements is
numerically time consuming.  Thus, we inject the transit after
generating the light curve.  This procedure has the potential to
systematically reduce or even eliminate the transit signal, because
generating the light curve and applying a seeing decorrelation tend to
``flatten'' a light curve.  To quantify the significance of this
effect, we inject transits in the raw photometric measurements before
the light curve generation procedure on several stars in the sample.
Comparing the detection probability obtained by injecting transits
before light curve generation to the detection probability obtained by
injecting the transit after light curve generation reveals that injecting
the transit after generating the light curve overestimates the
detection probability by $\sim 0.03$.  We decrease the calculated
probability at fixed period by 0.03 to account for this systematic
effect.

The 0.03 systematic overestimate in the detection probability
becomes increasingly important for correctly characterizing the
detection probability at long orbital periods.  For instance, the
detection probability for a star of median brightness
will be overestimated by $>$15\% for orbital periods $>4.0$ day and
1.5 $R_{J}$ companions if this systematic effect is not taken into
account.  The detection probability is overestimated by $>$50\% for
orbital periods $>$8.0 day without correction.  The results for 1.0
$R_{J}$ companions are even more severe.  The detection probability
would be overestimated by 50\% for periods beyond 1.8 day for a star
of median brightness without correction.

Based on the CMD of NGC 1245 \citep{BUR04}, this study contains light
curves for $\sim$ 2700 stars consistent with cluster membership.
Initially, we calculate the detection probability for 2 possible
companion radii: 1.0 and 1.5 $R_{J}$.  For each star, on average we
inject 50000 transits for a single companion radius at 150 different
semimajor axes.  In total, we inject and attempt to recover $\sim
2.7\times 10^{8}$ transits.  Current processors allow injection and
attempted recovery on order of 1 s per transit.  A single processor
requires $\sim$3000 days for the entire calculation.  Fortunately, the
complete independence of a transit injection and recovery trial allows
parallelization of the calculation.  We accomplish a parallel
calculation via a server and client architecture.  A server injects a
transit in the current light curve and sends it to a client for
recovery.

Based on the computing resources available, we employ two different
methods for communication between the server and clients.  Using a
TCP/IP UNIX socket implementation for communication between the server
and clients allows access to $\sim$40 single-processor personal
workstations connected via a local area network within the department
of astronomy at The Ohio State University.  Additionally, the
department of astronomy at The Ohio State University has exclusive
access to a 48 processor Beowulf cluster via the Cluster Ohio program
run by the Ohio Supercomputer Center.  The Message Passing Interface
(MPI) libraries provide communication between the server and clients
on the Beowulf cluster.  A Beowoulf cluster belonging to the Korean
Astronomy Observatory also provided computing resources for this
calculation.  C programming source code for either client-server
communication implementation is available upon request from the
author.

\begin{figure}
\plotone{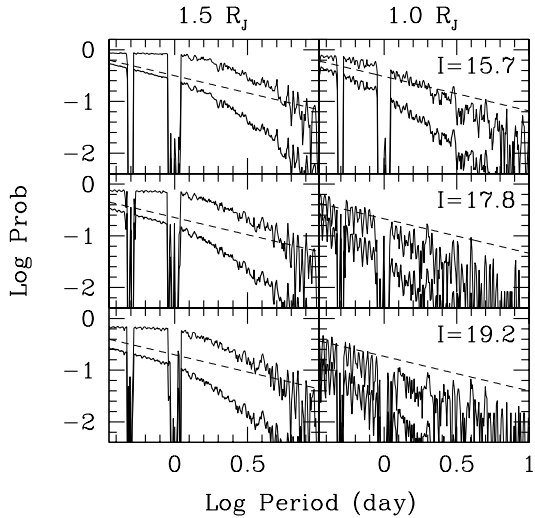}

\caption{Detection probability as a function of the orbital period 
is shown as the {\it heavy solid line}.  This is a product of the probability for a
transit to occur ({\it dash line}) and the probability that an
injected transit meets the selection criteria ({\it light solid
line}).  The panels from top to bottom show representative stars in
order of increasing apparent magnitude.  The {\it left} panels give
results for a 1.5 $R_{J}$ companion.  The {\it right} panels give
results for a 1.0 $R_{J}$ companion.\label{efffig}}

\end{figure}

The light solid line in Figure~\ref{efffig} shows the detection
probability, $P_{\epsilon}(a,R_{p})$, for three representative stars
in order of increasing apparent magnitude from top to bottom and for
the 2 companion radii, 1.0 and 1.5 $R_{J}$, on the left and right,
respectively.  In general, the probability nears 100\% completion for
orbital periods $\la 1.0$ day and then has a power law fall off toward
longer orbital periods.  The falloff in the detection probability toward longer orbital periods
partially results from the requirement of observing more than one transit.
The large drop in the detection probability
around 0.5 and 1.0 day orbital periods results from the selection
criteria we impose.  The narrow, non-zero spikes in the detection
probability near the 0.5 and 1.0 day orbital periods result from
injecting a transit at this period, but the BLS method returns a
best-fit period typically at the $\sim$0.66 day alias.

Figure~\ref{efffig} shows the detection probability with 3.3 times
higher resolution in orbital period and a lower, 1\%, error in the
detection probability at fixed orbital period than the actual
calculation.  Thus, the figure resolves variability in the detection
probability as a function of orbital period for probabilities
$\ga$1\%.  However, such fine details have negligible impact on the
results.

\subsection{Calculating $P_{T,i}(a,R_{p})$}

The probability for a transit to occur is $P_{T}=(R_{\star}+R_{p})/a$.
This transit probability assumes the transit is equally detectable for
the entire possible range of orbital inclinations that geometrically
result in a transit.  As $\cos i$ for the orbit approaches
$(R_{\star}+R_{p})/a$ the transit length and depth decreases,
degrading the transit S/N.  We address this when computing $P_{\epsilon}$
by injecting the transit with an even distribution in $\cos i$ between
the geometric limits for a transit to occur.  Thus, $P_{T}$ represents
the overall probability for a transit with high enough inclination to
begin imparting a transit signal, while the detailed variation of the
light curve signal for varying inclination takes place when
calculating $P_{\epsilon}$.  $P_{T}$ is shown as the dashed light line
in Figure~\ref{efffig}.  The heavy solid line in Figure~\ref{efffig}
is the product of $P_{\epsilon}$ and $P_{T}$.

\subsection{Calculating $P_{\rm mem}$}

The Monte Carlo calculation requires knowledge of the stellar
properties, and the given properties are only valid if the star is in
fact a bona fide cluster member.  An estimate of the field-star
contamination from the CMD provides only a statistical estimate of the
cluster membership probability.  Based on the study of the mass
function and field contamination in \citet{BUR04}, we estimate the
cluster membership probability, $P_{\rm mem}$, as a function of
stellar mass.  In brief, we start with a subsample of stars based on
their proximity to the best-fit cluster isochrone (selection on
$\chi^{2}_{\rm mem}<0.04$, see \S\ref{trncands}).  This sample
contains $N_{\star}\sim 2700$ potential cluster members, and the heavy
points in Figure~\ref{cmd} mark this cluster sample in the CMD.  The
best-fit isochrone allows an estimate of the stellar mass for each
member of the cluster sample, and we separate the sample into mass
bins.  Repeating this procedure on the outskirts of the observed field
of view, scaled for the relative areas, provides an estimate of the
field-star contamination in a given mass bin.  We fit $P_{\rm mem}$,
given in discrete mass bins, with a smooth spline fit for
interpolation.

\begin{figure}
\epsscale{1.3}
\plotone{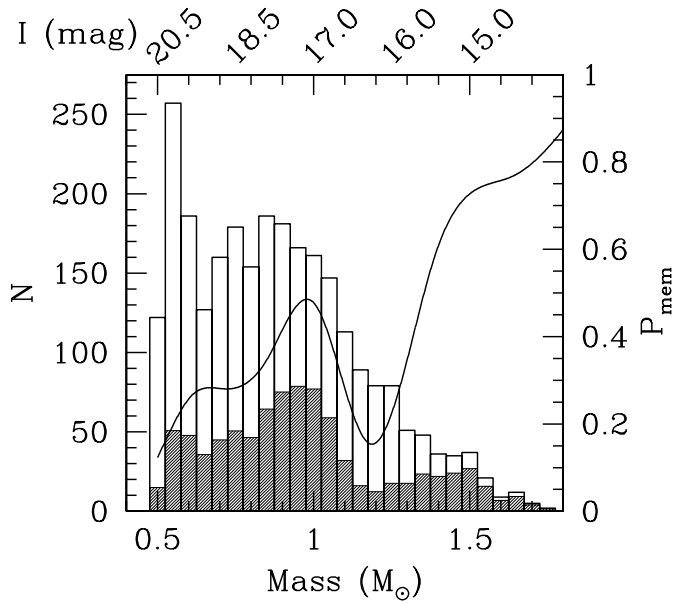}

\caption{Distribution of the potential cluster members as a function
of stellar mass ({\it open histogram}). The {\it solid line} shows the
membership probability (right hand ordinate) as a function of stellar
mass.  The {\it shaded histogram} shows the product of the potential
cluster member histogram and the cluster membership probability.  The
corresponding apparent $I$-band magnitude is given along the
top.\label{memprob}}

\end{figure}

The solid line in Figure~\ref{memprob} shows $P_{\rm mem}$ as a
function of stellar mass.  The corresponding probability is given on
the right side ordinate.  The clear histogram shows the distribution
of the potential cluster members as a function of mass.  The lower
shaded histogram shows the product of the potential cluster members
histogram and $P_{\rm mem}$.  This results in
effectively $N_{\star ,\rm eff}\sim$870 cluster members in total.  For
reference, the corresponding apparent $I$-band magnitude is given along
the top.

\section{Results}\label{results}

\subsection{Results Assuming a Power-law Orbital-period Distribution}

The previous section describes the procedure for calculating the
sensitivity of the survey to detect planetary companions as a function
of semimajor axis.  The results from this calculation enable us to
place an upper limit on the fraction of cluster members harboring
close-in companions given the null result.  However, calculating the
upper limit over a range of orbital periods necessitates assuming a
distribution of orbital periods for the planetary companions.
Radial velocity surveys characterize the distribution of extrasolar
planets in period as $dn\propto P^{-\gamma} dP$, with $0.7\la \gamma
\la 1.0$, corresponding to $dn\propto a^{-\beta} da$, with $0.5\la
\beta \la 1.0$ \citep{STE01,TAB02}.  These studies fit the entire
range of orbital periods ranging from several days to several years.
More recently, after an increase in the number of extrasolar planet
discoveries, \citet{UDR03} confirm a shortage of planets with $10\la P
\la 100$ day orbits.  Thus, the period distribution may take on
different values of $\gamma$ in the $P\la 10$ day and $P\ga 100$ day
regimes.

The initial extrasolar planet discoveries via the transit technique
had periods less than 3.0 days \citep{KON04}.  The detection of these
``Very Hot Jupiters,'' contrasted with the
results from radial-velocity surveys, which demonstrated a clear
paucity of planets with $P\la 3.0$ days.  After accounting for the
strong decrease in sensitivity of field transit surveys with increasing
period, \citet{GAU05A} demonstrated the consistency between the apparent lack
of VHJ companions in the radial velocity surveys and their discovery
in transit surveys.  They further demonstrated that VHJ appear
to be intrinsically much rarer than HJ (with 
$3 \leq P/\days \leq 9$.  We will therefore treat
VHJ as HJ distinct populations. 

Due to the incomplete knowledge of the actual
period distribution of extrasolar planets and its possible dependence
on the properties of the parent star, we provide upper limits assuming
an even logarithmic distribution of semimajor axis.  
Thus, we assume a
form of the joint probability distribution of the semimajor axis and
$R_{p}$ given by
\begin{equation}
\frac{d^{2}p}{dR_{p} da}=k\delta(R_{p}-R_{p}') a^{-1}\label{uplimint},
\end{equation}
where $k$ is the normalization constant, $\delta$ is the Dirac delta
function, and $R_{p}'$ is the planet radius.  We initially give results
for $R_{p}'=1.0$ and $1.5$ $R_{J}$.   We follow \citet{GAU05A} and show
results for HJ (3.0$<$P$<$9.0 day) and VHJ (1.0$<$P$<$3.0 day) ranges.
In addition, we show results for a more extreme population of
companions with $P_{\rm Roche}<P<1.0$ day, where $P_{\rm Roche}$ is
the orbital period at the Roche separation limit, which we designate
as Extremely Hot Jupiter (EHJ).  Assuming a negligible companion mass,
the Roche period depends solely on the density of the companion.
Jupiter, Uranus, and Neptune have nearly the same $P_{\rm Roche}\sim
0.16$ day.  

\begin{figure}
\epsscale{1.2}
\plotone{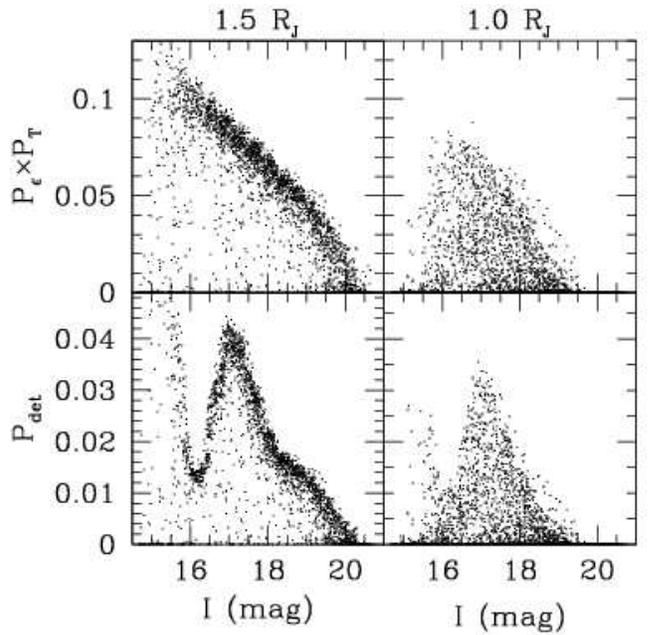}

\caption{Probability for transit detection as a function of the
apparent $I$-band magnitude assuming an even logarithmic distribution
in semimajor axis from 1.0$<P<$3.0 day.  The {\it top} panels assume
$P_{\rm mem}=1.0$.  The {\it left} panels show results for a 1.5
$R_{J}$ companion.  The {\it right} panels show results for a 1.0
$R_{J}$ companion.  The {\it bottom} panels are the same as the top
panels, but they take into account the membership probability $P_{\rm mem}$.\label{powerlawfig}}

\end{figure}

Figure~\ref{powerlawfig} shows the probability for detecting a
VHJ (1.0 day $\leq P \leq$ 3.0 day) companion with an even logarithmic
distribution in semimajor axis as a function of apparent $I$-band
magnitude.  The left and right panels show results for a 1.5 and 1.0
$R_{J}$ companion, respectively.  The top panels of
Figure~\ref{powerlawfig} show the probability for detecting an
extrasolar planet, $P_{\rm det}$, assuming $P_{\rm mem}=1.0$.  The
bottom panels show $P_{\rm det}$ after taking into account $P_{\rm
mem}$.  The results for 1.0 $R_{J}$ companions broadly scatter across
the full range of detection probability.  However, the 1.5 $R_{J}$
companion results delineate a tight sequence in detection probability
as a function of apparent magnitude.

The 1.5 $R_{J}$ companion signal lies many times above the rms scatter
in the light curve (see Figure~\ref{magrms}).  Thus, a single
measurement contributes a large fraction of the S/N required for
detection.  In this limit, the observing window function mainly
determines the detection probability, and as we show in
\S\ref{thyeffdisc} the result is similar to results obtained by the
theoretical detection probability framework of \citet{GAU00}.
However, the 1.0 $R_{J}$ companion transit signal comes closer to the
detection threshold.  \citet{PEP05} describe the sensitivity of a
transit survey as a function of planet radius.  The sensitivity of a
transit survey depends weakly on $R_{p}$ until a critical radius is
reached when the S/N of the transit falls rapidly.  The sensitivity of
the survey for 1.0 $R_{J}$ is near this threshold, hence the large
scatter in the detection probability.

With the detection probabilities for all stars in the survey for the
assumed semimajor axis distribution, we can calculate the expected
number of detections scaled by the fraction of cluster members with
planets.  Thus, from the Poisson distribution, a null result 
is inconsistent at the $\sim$95\% level when $N_{\rm
det}\sim 3$.  This allows us to solve for the 95\% confidence upper
limit on the fraction of cluster members with planets using Eq.\ \ref{eqn:ndet}.
This gives,
\begin{equation}
f_{\star} \le 3.0/\sum_{i=1}^{N_{\star}}P_{\rm det,i}\qquad {\rm (95\%~c.l.)}.\label{uplimeq}
\end{equation}

\begin{figure}
\plotone{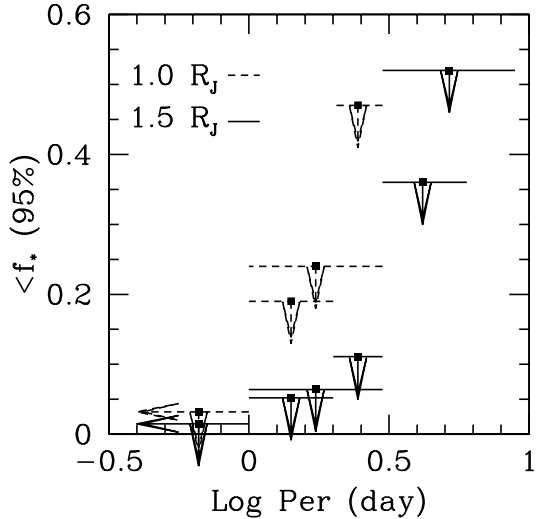}

\caption{Upper limit (95\% Confidence) on the fraction of stars in the
cluster with companions for several ranges in orbital period assuming
an even logarithmic distribution in semimajor axis.  The {\it solid
lines} show results for a 1.5 $R_{J}$ companion.  The {\it dash lines}
show results for a 1.0 $R_{J}$ companion.\label{uplimit}}

\end{figure}

Figure~\ref{uplimit} shows the 95\% confidence upper limit on the
fraction of stars with planets in NGC 1245 for several ranges of
orbital period.  The solid and dashed lines give results for 1.5 and 1.0
$R_{J}$ companions, respectively.  For 1.5 $R_{J}$ companions we limit
the fraction of cluster members with companions to $<$1.5\%, $<$6.4\%,
and $<$52\% for EHJ, VHJ, and HJ companions, respectively.  For 1.0
$R_{J}$ companions, we find $<$2.3\% and $<$15\% have EHJ and VHJ
companions, respectively.

The detection probability decreases rapidly with orbital period beyond
1.0 day.  As a result, the survey does not reach the sensitivity
needed to place an interesting upper limit on 1.0 $R_{J}$ companions beyond $P>3.0$
day.  

We further divide the VHJ period range and show upper limits for
the period ranges 
1.0< $P/\days$ <2.0 and 2.0$<P/\days<$3.0, which we denote as $P_{12}$ and $P_{23}$.  
For 1.5 $R_{J}$ companions we limit $f_{\star}$ to $<$5.2\%
and $<$11\% for $P_{12}$ and $P_{23}$, respectively.  For 1.0 $R_{J}$
companions we limit $f_{\star}$ to $<$19\% and $<$47\% for $P_{12}$
and $P_{23}$, respectively.  We also divide the HJ period range and
limit $f_{\star}$ for 1.5 $R_{J}$ companions in the 3.0$<P/\days<$6.0 to
$<$36\%.

\subsection{Results for Other Companion Radii\label{uplimradsec}}

\begin{figure}
\plotone{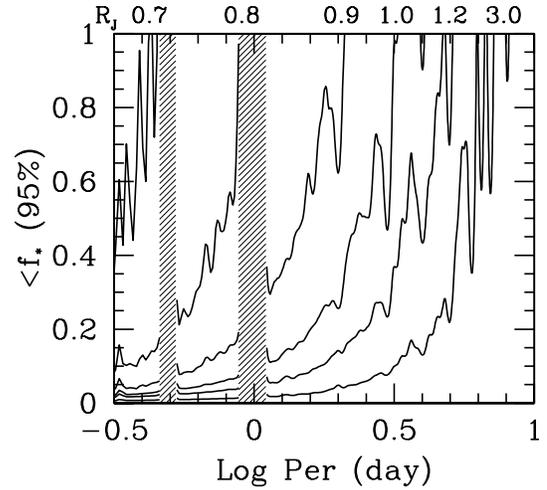}

\caption{Upper limit (95\% Confidence) on the fraction of stars in the
cluster with companions for several companion radii as label along the
top.  The result for a 1.0 $R_{J}$ companion is based on the entire
sample, whereas the results for the other companion radii are based on
a subsample of $N_{\star}=100$ stars.  The shaded regions denote orbital periods
removed by the selection criteria in order to eliminate false-positive
transit detections that occur around the diurnal period and 0.5 day
alias.\label{uplimrad}}

\end{figure}

Due to computing limitations we calculate detection probabilities for
the entire cluster sample only for 1.5 and 1.0 $R_{J}$ companions.  In
\S\ref{uplimiterrsec} we show that an upper limit determination using
a subsample of the stars with size $N_{\star ,\rm SS}=100$
approximates the results based on the entire stellar sample.  Thus, we
calculate upper limits for a variety of companion radii using
$N_{\star ,\rm SS}=100$ randomly chosen stars in the sample.  Instead
of showing upper limit results over a range of orbital periods, we
derive upper limits at fixed period by replacing the semimajor axis
distribution with a $\delta(a-a_{o})$ function in
Equation~\ref{uplimint}.  To obtain results at fixed period, each star
has a different $a_{o}$ that depends on the stellar mass.
Figure~\ref{uplimrad} shows the upper limit on the fraction of stars
with planets in the survey as a function of orbital period.  The lines
show results for various values of the companion radius in terms of
$R_{J}$ as indicated by the label next to each line along the top of
the figure.  The shaded regions denote orbital periods removed by the
selection criteria in order to eliminate false-positive transit
detections that occur around the diurnal period and 0.5 day alias.  At
smaller companion radii, the transit S/N$\propto R_{p}^{2}$ drops
quickly.  Toward larger companion radii the S/N of the transit
saturates and the observational window function increasingly dominates
the survey effectiveness.  The survey cannot detect companions with
$R_{p}>3.5 R_{J}$ as the transit/eclipse becomes too deep given the
removal of measurements that deviate by more than 0.5 mag from the
mean light-curve level.

\section{Error in the Upper Limit}\label{uplimiterrsec}

In this section we discuss several sources of error present when
determining an upper limit on the fraction of stars with planets.

\subsection{Error When Using a Subsample}

Computing power limitations discourage calculating detection
probabilities over the entire cluster sample.  Thus, we first
characterize the error associated with determining an upper limit
using only a subset of the entire cluster sample.  Starting with
Equation~\ref{uplimeq}, we derive an error estimate when using a
subsample by the following means.  Replacing the summation over
$P_{i,\rm det}$ with the arithmetic mean, $\langle P_{\rm det} \rangle$,
Equation~\ref{uplimeq} becomes
\begin{equation}
f_{\star}=3.0/(N_{\star}\langle P_{\rm det} \rangle)\label{uplimave}.
\end{equation}
By propagation of errors, the error in the upper limit is given by
\begin{equation}
\sigma_{f}=\frac{3.0}{N_{\star}}\frac{\sigma_{\langle P \rangle}}{\langle P_{\rm det} \rangle^{2}},\label{uplimerreq}
\end{equation}
where $\sigma_{\langle P \rangle}$ is the error in the mean detection
probability.  The error in the mean detection probability scales as
$\sigma_{\langle P \rangle}=\sigma_{P}/\sqrt{N_{\star ,\rm SS}}$, where
$\sigma_{P}$ is the intrinsic standard deviation of the distribution of
$P_{i,\rm det}$ values and $N_{\star ,\rm SS}$ is the size of the subsample.

\begin{figure}
\plotone{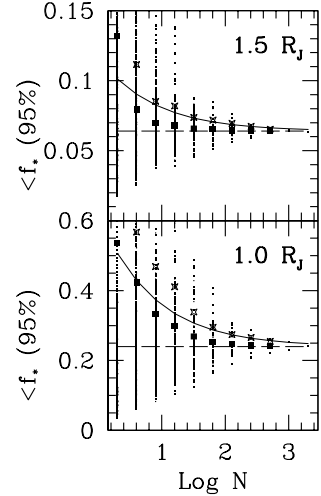}

\caption{Estimates for the upper limit (95\% Confidence) on
the fraction of stars in the cluster as a function of the sample size
employed in making the estimate are shown as {\it small points}.
We have assumed an even logarithmic distribution in periods between
$1.0<P/\days<3.0$ orbital period.   The {\it dash
line} shows the upper limit based on the entire sample.  The average
upper limit at fixed sample size is given by {\it square points}.  The
sequence of {\it open stars} gives the standard deviation in the
distribution of upper limits at fixed sample size.  The {\it solid
line} shows the error model estimate for the standard deviation in the
upper limit.  The {\it top} panel gives results for a 1.5 $R_{J}$ companion, and the {\it bottom} panel gives results for a 1.0 $R_{J}$
companion.\label{uplimiterr}}

\end{figure}

We empirically test this error estimate by calculating the upper limit
with subsamples of increasing size.  The small points in
Figure~\ref{uplimiterr} show the upper limit on the fraction of stars
with planets as a function of the subsample size.  The upper limit
calculation assumes an even logarithmic distribution of semimajor axis
for companions with $1.0\leq P\leq 3.0$ day for 1.5 and 1.0 $R_{J}$
radius companions, top and bottom panels, respectively.  Neighboring
columns of upper limits differ by a factor of 2 in the subsample size.
We randomly draw stars from the full sample without replacement,
making each upper limit at fixed sample size independent of the
others.  The dashed line represents the upper limit based on the full
cluster sample.

The distribution of upper limits around the actual value possesses a
significant tail toward higher values.  This tail results from the
significant number of stars with $P_{\rm tot}=0.0$.  At fixed sample
size, the large square point represents the mean upper limit.
Using subsamples sizes of $N_{\star ,\rm SS}\lesssim 20$ tends
to systematically overestimates the true upper limit.  The open star symbol
represents the $1-\sigma$ standard deviation of the distribution at
fixed sample size.  The solid line shows the error estimate from
Equation~\ref{uplimerreq}.  Despite the non-Gaussian nature of the
underlying distribution, the error estimate in the upper limit
roughly corresponds with its empirical determination especially toward
increasing $N_{\star ,\rm SS}$ where the systematic effects become
negligible.  From Figure~\ref{uplimiterr}, we conclude that 
adopting $N_{\star ,\rm SS}\ga 100$
provides adequate control of the random and systematic errors in
calculating an upper limit, without becoming numerically prohibitive.
This verifies the procedure for estimating the upper limit for a
variety of companion radii in \S\ref{uplimradsec}.

\subsection{Error in Determining Sample Size}

Up to this point, we have mainly addressed sources of error directly
associated with determining $P_{\epsilon}$.  However, the upper limit
error budget contains an additional source of error from uncertainties
in determining $P_{\rm mem}$.  This additional source of error
directly relates to the accuracy in determining the number of single
main-sequence stars in the survey.

We characterize this error as follows.  At fixed orbital period,
$\langle P_{\rm det} \rangle=\langle P_{\rm mem} P_{\epsilon}
P_{T} \rangle$.  Given that  $P_{\rm mem}$ is nearly independent of the other
terms, the previous average is separable, such that $\langle P_{\rm det} \rangle=\langle P_{\rm mem} \rangle \langle P_{\epsilon} P_{T} \rangle $.  This
separation changes the derived upper limit by a negligible 0.3\%
relative error.  The separation allows us to rewrite
Equation~\ref{uplimave} as
\begin{equation}
f_{<,95}=3.0/(N_{\star ,\rm eff}\langle P_{\epsilon} P_{T} \rangle),
\end{equation}
where $N_{\star ,\rm eff}=N_{\star}\langle P_{\rm mem} \rangle$ is
the effective number of cluster members in the sample after taking
into account background contamination.  Thus, $N_{\star ,\rm eff}$
carries equal weight with $\langle P_{\epsilon} P_{T} \rangle$ in the upper-limit
error budget.


The ability to determine $N_{\star ,\rm eff}$ accurately provides an
advantage for transit surveys toward a rich stellar cluster rather
than toward a random galactic field.  Even though methods based on the
cluster CMD statistically determine cluster membership, they
concentrate on a narrow main-sequence region to search for planets
where the cluster counts significantly outweigh the background
contamination counts.  By concentrating on the main sequence of a
cluster, this survey has only $\sim 68\%$ contamination by background
stars.  In contrast, random galaxy fields contain $\gtrsim 90\%$
contamination by subgiant and giant stars for V$<$11 surveys
\citep{GOU03}.  Overall, $N_{\star ,\rm eff}$ has an 8\% error, which
propagates to a relative error of 8\% in the upper limit.  The error
in $N_{\star ,\rm eff}$ comes from subtracting the star counts
observed within a 12.7$\arcmin$ radius of the cluster center by the
control field star counts outside this radius.  The error is larger
than the Poisson error of $N_{\star ,\rm eff}=870$ since the control
field star count is scaled to match the larger cluster field area.

\subsection{Error Due to Blends and Binaries\label{binstat}}

The final source of error we address results from stellar blends due
to physical binaries or chance, line-of-sight associations.  The
additional light from an unresolved blend dilutes a transit signal
from one component of the blend.  Thus, we overestimate the ability to
detect a transit around blends.  However, a compensatory effect arises
since the extra light from a blend results in an overestimate in the
stellar mass and radius, which in turn results in modeling a shallower
transit.  Modeling such details is not possible without knowing the
binary nature for each object, but we can estimate the number of stars
affected by assuming binary star statistics as measured in the field.
Due to low stellar crowding, we estimate chance blends have a
negligible effect in comparison to physically associated binaries
\citep{KIS05}.  Finding charts in Figure~\ref{find} demonstrate the
stellar crowding conditions of the survey.

The latest Coravel radial velocity survey dedicated to F7-K field
dwarfs \citep{HAL04} and the visual binary and common proper motion
pairs survey of \citet{EGG04} provide the basis for the binary star
estimates.  Overall they find a binary frequency of 56\% for systems
with $\log (P/\days)\leq 6.31$.  However, due to the strong dependence of
luminosity on the stellar mass only systems with mass ratio, $q>0.6$,
significantly contribute light to dilute the transit signal.  For
lower mass ratios the lower mass component contributes $<20\%$ of the
total system flux.  When taking binaries across the entire range of
orbital periods the mass-ratio distribution peaks near $q\sim 0.2$ and
slowly drops toward higher q \citep{DUQ91}.  From Figure 10 in
\citet{DUQ91}, only $\sim 20\%$ of their binary systems have $q>0.6$.
Thus, if the binary statistics for the cluster matches the field
dwarfs, transit dilution occurs for $\sim 11\%$ of the stellar sample.
The radial velocity survey for binaries in the Pleiades and Praesepe
clusters reveals consistency with the frequency of binaries in the
field surveys \citep{HAL04}.

In principle, the data from this survey can also answer whether the
binary statistics of the cluster matches the field dwarfs.  However,
the statistical methods and selection criteria described in this study
do not optimally detect interacting and eclipsing binaries.
Additionally, in order to reach planetary companion sensitivities, we
remove light-curve deviations beyond 0.5 mag as discrepant, which
removes the deep eclipses.  

\subsection{Overall Error}

The errors involved with determining the number of cluster members
dominates the error budget in determining the upper limit.  However,
as discussed in \S\ref{effcalc}, this is only true if one quantifies
and corrects for the systematic overestimate in detection probability
due to a reduction in the transit signal from the procedures of
generating and correcting the light curve.  For instance, at the
median stellar brightness for this survey, the detection probability is
overestimated by $>$15\% for orbital periods $>$4.0 day and $>$1.0 day
for 1.5 and 1.0 $R_{J}$ companions, respectively, without correction.
Since we characterize this systematic effect, the error
in determining the number of cluster members dominates the error
budget.

Additionally, the potential for a large contamination of binaries
diluting the transit signal necessitates an asymmetrical error bar.
We roughly quantify the error estimate resulting from binary
contamination from the field dwarf binary statistics.  From the arguments
in the previous section, we adopt 11\% as a $1-\sigma$ systematic
fractional error due to binary star contamination.  Overall, combining
this systematic error with the 7\% fractional error in determining the
cluster membership, upper limits derived from the full stellar sample
contain a $^{+13\%}_{-7\%}$ fractional error.

\section{Discussion}\label{discussion}

Along with this work, several other
transit surveys have quantified their detection probability from actual
observations in an attempt to constrain the fraction of stars with
planets or quantify the consistency with the solar neighborhood radial
velocity planet discoveries \citep{GIL00,WEL05,MOC05,HID05,HOO05}.
Unfortunately, a direct comparison of upper limits from this work with these
other transit surveys cannot be made.  Until this study, none of
the previous studies have quantified the random or systematic errors
present in their techniques in sufficient detail to warrant a
comparison.  Additionally, previous studies do not 
have quantifiable
selection criteria that completely eliminate false-positive transit
detections due to systematic errors in the light curve, a necessary
component of an automated Monte Carlo calculation.

\subsection{Initial Expectations vs. Actual Results}

In the meantime, we can discuss why the initial estimate of finding
two planets assuming 1\% of stars have $R_{J}$ companions evenly
distributed logarithmically between 0.03 to 0.3 AU \citep{BUR03}
compares to the results from this study, which indicate that we
expected to detect 0.1 planets.  The initial estimates for the
detection rate are based on the theoretical framework of
\citet{GAU00}.  Given a photometric noise model, observational window,
and S/N of the transit selection criteria, the theoretical framework
yields an estimate of the survey detection probability.  This
theoretical detection probability coupled with a luminosity function
for the cluster determines the expected number of detections.  As we
show next, the initial estimates did not account for the light curve
noise floor or detector saturation, and contain optimistic estimates for
the sky background and luminosity function.  In addition, the initial
estimates could not have accounted for the 50\% reduction in signal
for the majority of the light curves due to the detector error
discussed in \S\ref{sec:noise}.  Finally, as discussed in detail by
\citet{PEP05} and demonstrated explicitly here, the detection
probability is very sensitive to the precise error properties near the
critical threshold of detection, which for this survey is just reached
for $R_J$ companions.

\begin{figure}
\epsscale{1.3}
\plotone{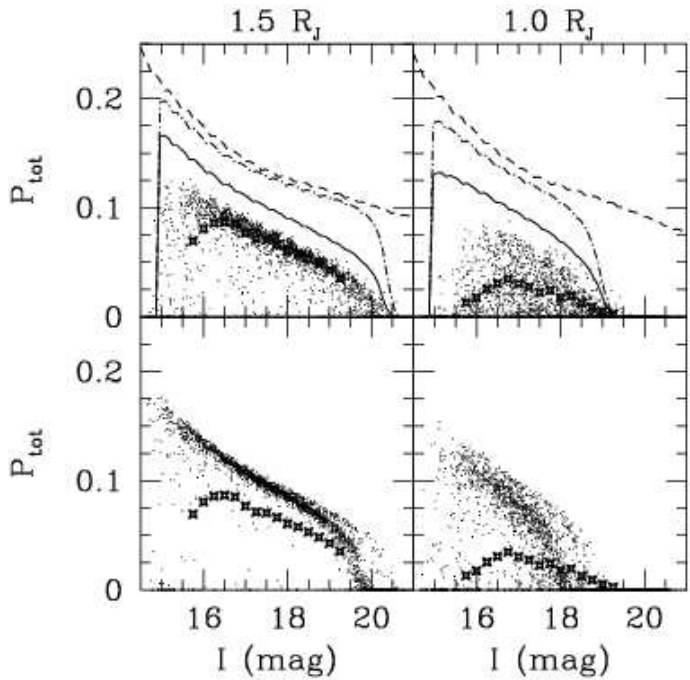}

\caption{{\it Top}: Probability for transit detection as a function of
the apparent $I$-band magnitude assuming an even logarithmic
distribution in semimajor axis from 1.0$<P<$3.0 day and $P_{\rm
mem}=1.0$ using the Monte Carlo calculation of this study({\it small
points}).  The binned average of the Monte Carlo results is denoted by
{\it open stars}.  The {\it dash line} shows the expected probability
for transit detection based on a theoretical calculation prior to this
survey.  The {\it dot dash line} shows the theoretical probability for
transit detection assuming a photometric noise model appropriate for
the survey.  The {\it solid line} shows the theoretical probability
for transit detection with an accurate photometric noise model for the
survey and including the effects of limb darkening.  The {\it left}
panel shows 1.5 $R_{J}$ companion results.  The {\it right} panel
shows 1.0 $R_{J}$ companion results.  {\it Bottom}: Shows the
theoretical probability for transit detection allowing each star of
the survey to have its empirically determined photometric noise and
including the effects of limb darkening ({\it small points}).  The
open stars are reproduced from the top panels.\label{effthy}}

\end{figure}

The top panels of Figure~\ref{effthy} compare the detection
probability of the Monte Carlo calculation of this study to the
initial theoretical estimate.  The small points replicate the
Monte Carlo results from the top panels of Figure~\ref{powerlawfig},
while the dashed line shows the detection probability based on the
initial theoretical expectations.  The initial theoretical
expectations clearly overestimate the detection probability.  The
bright end continues to rise due to ignoring the effects of detector
saturation and the photometric noise floor.  The faint end does not
cutoff due to an underestimated sky brightness.  The initial estimate
of the sky brightness, 19.5 mag arcsec$^{-2}$, compares optimistically
to the range of sky brightnesses encountered during the actual
observations.  The sky varied between 17.5 and 19.0 mag arcsec$^{-2}$
over the course of the observations.  The full lunar phase took place
near the middle of the observation, and the Moon came within 40$\degr$
of the cluster when nearly full.

The initial estimate for the cluster
luminosity function simply selected cluster members via tracing by eye lines
that bracket the main sequence in the CMD.  This crude
technique led to an estimated 3200 cluster members down to $I\sim$20.
A careful accounting of the field star contamination results in only
$\sim$870 cluster members in the survey.  The luminosity function
overestimate and the expected sensitivity to transits around
the bright and faint cluster members leads to a factor of 4-5
overestimate in the number of cluster members in the survey.
Additionally, the factor of 4-5 overestimate of the initial detection
probability when compared to binned average detection probability for
the Monte Carlo results (open stars in Figure~\ref{effthy}), easily
accounts for the factor of 20 difference in the overall number of expected
detections (for $R=R_J$).

\subsection{Improving Theoretical Expectations\label{thyeffdisc}}

Clearly, accurate and realistic transit detection statistics requires
more detailed analysis than these early estimates and more careful
theoretical work has already been done \citep{PEP05}.  In the case of
an open cluster, delineating cluster membership by tracing the main
sequence in the CMD overestimates the number of cluster members.  A
careful subtraction of the field contamination is necessary in order
to extract an accurate cluster-member count. 

A photometric noise model that accurately reflects the quality of
observations is the next step in correctly calculating a theoretical
detection probability.  From Figure~\ref{magrms}, we estimate the
actual photometric noise present in the data.  This includes the
proper sky measurement and systematic floor in the photometric
precision.  With a noise model similar to the lower solid line in
Figure~\ref{magrms}, we recalculate the theoretical detection
probability.  The dot dash line in Figure~\ref{effthy} shows the
resulting detection probability still overestimates the Monte Carlo
results.  However, it does agree with the faint-end cutoff of the
Monte Carlo calculation.  We impose the bright-end cutoff due to
saturation effects at the same magnitude as the observed increase in
light curve rms as shown in Figure~\ref{magrms}.

For these results we include an additional effect not taken into
account by \citet{GAU00}.  We multiply the transit S/N selection
criteria, Equation 5 of \citet{GAU00}, by $\sqrt{{\rm max}(N_{\rm
obs},1.7)}$, where $N_{\rm obs}$ is the typical number of transits
detected throughout the observing run.  The $N_{\rm obs}=1.7$ floor in
this factor corresponds to the requirement of observing the transit
twice multiplied by the observing efficiency.  For simplicity, we take
$N_{\rm obs}=N_{\rm tot}/P\times 0.2$, where $N_{\rm tot}=16$, the
length of the observing run in days, and the factor of 0.2 accounts
for the actual observational coverage encountered during the run.

Given that the theoretical calculation still overestimates the Monte Carlo
results, to increase the realism of the theoretical detection
probability, we include a linear limb-darkening law, which effectively
weakens the transit depth.  We solve for the factor $G$, Equation 6 of
\citet{GAU00}, assuming a linear limb-darkening parameter, $\mu=0.6$,
for all stars.  The inclusion of limb darkening significantly impacts
the theoretical detection probability as the solid line in
Figure~\ref{effthy} demonstrates.  Although the theoretical detection
probability still overestimates the upper envelope of results from the
Monte Carlo calculation, the level of agreement, after including an
accurate photometric noise model and limb darkening, shows significant
improvement over the initial estimates.

Despite the improved agreement, the Monte Carlo detection probability
calculation shows significant scatter at fixed magnitude.  The
theoretical probability treats all stars at fixed
magnitude as having the same noise properties.  With the theoretical
detection probability we can address whether the scatter in detection
probability at fixed magnitudes results from the observed scatter in
noise properties at fixed magnitude as shown in Figure~\ref{magrms}.
Thus, we calculate a theoretical detection probability for each star
individually using the measured rms in the light curve for each star
to determine the theoretical transit S/N selection criteria using Equation 5 of
\citet{GAU00}.  The small points in the bottom panels of
Figure~\ref{effthy} show the resulting theoretical detection probability.

Some of the scatter in detection probability results from the scatter
in noise properties as a function of magnitude.  The heavy star points
represent the average Monte Carlo detection probability in 0.25
magnitude bins.  In the case of the 1.5 $R_{J}$ companions, the signal
is large in comparison to the photometric noise.  The left panels of
Figure~\ref{effthy} demonstrate the theoretical detection probability
overestimates the Monte Carlo detection probability by only 20\%.
However, the closer the transit signal approaches the systematic and
rms noise, the theoretical detection probability strongly
overestimates the actual detection probability.  In the case of 1.0
$R_{J}$ companions (right panels of Figure~\ref{effthy}), the
theoretical calculation overestimates the Monte Carlo results by 80\%.
Thus, we urge caution when relying on a theoretical detection
probability when the survey is near the critical threshold for transit
detection.  Such is the case for 1.0 $R_{J}$ companions in this survey.

\subsection{Planning Future Surveys}

Even though the theoretical calculation overestimates the absolute detection
probability by a factor of $<$2, tests on a small sample of stars
with the Monte Carlo calculation reveal it provides a much higher
relative accuracy.  Thus, the computationally efficient theoretical
calculation allows us to examine the relative change in the detection
probability for a given change in survey parameters.  For planning
future surveys it is essential to decide between increasing the number
of stars by observing another cluster or improving the detection
probability by increasing the length of observations on a single
cluster.  As shown in \S\ref{results}, the upper limit scales linearly
with the sample size, thus keeping everything else constant,
increasing the sample size by a factor of 2 improves the
upper limit by a factor of 2.

Using the theoretical detection probability framework, we can quantify
the improvement in sensitivity for a survey twice as long.  We assume
a survey twice as long consists of an observing window identical to
the current survey for the first half and repeats the observing window
of the current survey for the latter half.  The upper limit improves
only by a factor of 1.3 for a logarithmic distribution of VHJ planets.
However, the upper limits for HJs with 3.0 to 9.0 day orbital periods
decrease by a factor of 2.6.  Thus, not only is it more efficient to
observe this cluster twice as long, but the analysis of \citet{GAU05A}
reveals a 5-10 times larger HJ population than the the VHJ population.
This strongly suggests transit surveys with a single observing site
require month long runs for maximum efficiency in detecting HJ
companions.

Figure~\ref{effthy} reveals little improvement in the detection
probability occurs for increasing the photometric precision, at least
for 1.5 $R_{J}$ companions.  To first order, the photometric precision
determines the faint-end cutoff in the detection probability.  Thus, a
lower sky background or improved photometric precision predominately
effects the number of stars in the survey rather than the detection
probability.  However, improving the photometric precision does lead
to increasing the sensitivity for smaller radius companions.  In the
case of 1.0 $R_{J}$ companions, the rms in the light curve typically
is $\lesssim$1.8 times lower than the transit signal.  As shown in the
previous section, the theoretical detection probability breaks down
for such low precision.  In the case of 1.5 $R_{J}$ companions, the
rms in the light curve typically is $\lesssim$4 times lower than the
transit signal.  Thus, for the 1.0 $R_{J}$ results to reach the same
sensitivity as the 1.5 $R_{J}$ results, improvement in the light curve
rms is necessary until the transit S/N is above a critical threshold
when the detection probability is weakly dependent on $R_{p}$
\citep{PEP05}.

According to a recent review of radial velocity detected planets,
$1.2\pm 0.3\%$ of solar neighborhood stars have HJ companions
\citep{MAR05}.  This survey of NGC 1245 reached an upper limit of 52\%
of the stars having 1.5 $R_{J}$ HJ companions.  As mentioned
previously, a survey lasting twice as long can reduce this upper limit
to 21\%.  Reaching similar sensitivity as the radial velocity results
requires observing additional clusters in order to increase the number
of stars in the sample.  This survey has $\sim 870$ cluster members
and $\sim 740$ of them have nonzero detection probability for 1.5
$R_{J}$ VHJ companions.  Hence a total sample size of $\sim 7400$
dwarf stars observed for a month will be needed to help constrain the
fraction of stars with planets to a 2\% level (comparable to radial
velocity results).  Assuming that the observed HJ frequency of $\sim 1\%$ 
remains valid for a variety of stellar
environments, we expect to detect one planet every 5000 dwarf stars
observed for a month.  Results for 1.0 $R_{J}$ companions without
substantial improvement in the photometric precision likely will
require a small factor larger sample size.

\section{Conclusion}\label{conclusion}

In this study we complete the analysis of a 19-night search for
transiting extrasolar planets orbiting members of the open cluster NGC
1245.  An automated transit search algorithm with quantitative
selection criteria finds six transit candidates; none are bona
fide planetary transits.  Thus,
this work also details the procedure for analyzing the null-result
transit search in order to determine an upper limit on the fraction of
stars in the cluster harboring close-in $R_{J}$ companions.  In
addition, we outline a new differential photometry technique that
reduces the level of systematic errors in the light curve.

A reliable upper limit requires quantifiable transit selection
criteria that do not rely on visual, qualitative judgments of the
significance of a transit.  Thus, we develop completely quantitative
selection criteria that enable us to calculate the detection
probability of the survey via Monte Carlo techniques.  We inject
realistic limb-darkened transits in the light curves and attempt their
recovery.  For each star we inject 100,000 transits at a variety of
semimajor axes, orbital inclination angles, and transit phases,
to fully map the detection probability for 2700 light curves
consistent with cluster membership based on their position in the CMD.
After characterizing the field contamination, we conclude the sample
contains $\sim$870 cluster members.

When calculating a 95\% confidence upper limit on the fraction of
stars with planets, we assume companions have an even logarithmic
distribution in semimajor axis over several ranges of orbital period.
We adopt the period ranges as outlined by \citet{GAU05A}, for HJ and
VHJ companions, and an as of yet undetected population with P$<$1.0
day, which we denote as Extremely Hot Jupiters (EHJ).  For NGC 1245,
we limit the fraction of cluster members with 1.0 $R_{J}$ companions
to $<$3.2\% and $<$24\% for EHJ and VHJ companions, respectively.  We
do not reach the sensitivity to place any meaningful constraints on
1.0 $R_{J}$ HJ companions.  For 1.5 $R_{J}$ companions we limit the
fraction of cluster members with companions to $<$1.5\%, $<$6.4\%, and
$<$52\% for EHJ, VHJ, and HJ companions, respectively.

We also fully characterize the errors associated with calculating the
upper limit.  We find the overall error budget separates into two
equal contributions from error in the total number of single dwarf
cluster members in the sample and the error in the detection
probability.  After correcting the detection probability for
systematic overestimates that become increasingly important for
detecting transits toward longer orbital periods (see
\S\ref{effcalc}), we conclude that random and systematic errors in
determining the number single dwarf stars in the sample dominate the
error budget.  \S\ref{results} details the error analysis, and
overall, we assign a $^{+13\%}_{-7\%}$ fractional error in the
upper limits.

In planning future transit surveys, we demonstrate that observing NGC 1245
for twice as long will reduce the upper limits for the important HJ
period range more efficiently than observing an additional cluster of
similar richness as NGC 1245 for the same length of time as this data
set.  To reach a $\sim$ 2\% upper limit on the fraction of stars with
1.5 $R_{J}$ HJ companions, where radial velocity surveys currently measure
1.3\% \citep{MAR05}, we conclude a total sample size of $\sim 7400$
dwarf stars observed for a month will be needed.  If 1\% of stars have
1.5 $R_{J}$ HJ extrasolar planets, we expect to detect one planet
every 5000 dwarf stars observed for a month.  Results for 1.0 $R_{J}$
companions without substantial improvement in the photometric
precision likely will require a small factor larger sample size.

\acknowledgements This publication was not possible in a timely manner
without the gracious donation of computing resources by the following
individuals: D. An, N. Andronov, M. Bentz, E. Capriotti, J. Chaname,
G. Chen, X. Dai, F. Delahaye, K. Denney, M. Dietrich, S. Dong,
S. Dorsher, J. Escude, D. Fields, S. Frank, H. Ghosh, O. Gnedin,
A. Gould, D. Grupe, J. Guangfei, C. Onken, J. Marshall, S. Mathur,
C. Morgan, N. Morgan, S. Nahar, J. Pepper, B. Peterson, J. Pizagno,
S. Poindexter, J. Prieto, B. Ryden, A. Steed, D. Terndrup, J. Tinker,
D. Weinberg, R. Williams, B. Wing, J. Yoo.  We thank C. Han for the
donation of supercomputing resources belonging to the Korea Astronomy
Observatory and Astrophysical Research Center for the Structure and
Evolution of the Cosmos (ARCSEC) of Korea Science and Engineering
Foundation (KOSEF) through Science Research Program (SRC) program.
This publication makes use of supercomputer resources through the
Cluster Ohio Project Rev3, an initiative of the Ohio Supercomputer
Center, the Ohio Board of Regents, and the OSC Statewide Users Group.
This work was supported by NASA grant NAG5-13129 and a Menzel
Fellowship from the Harvard College Observatory.

\begin{figure}
\plotone{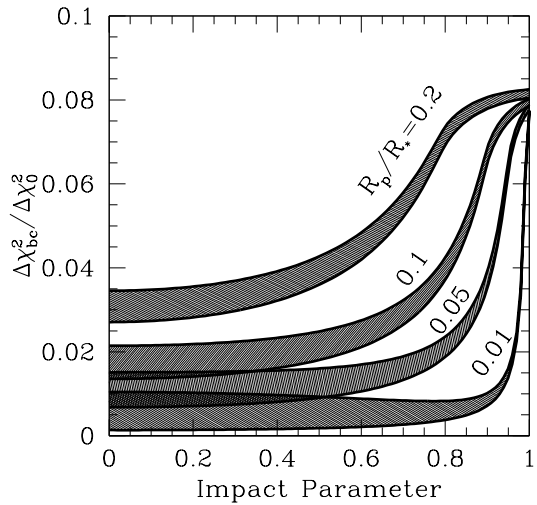}

\caption{The lines show $\Delta\chi^2_{bc}$, the difference in $\chi^2$ between a boxcar fit to a
planetary transit across a limb-darkened star and the exact model fit, 
normalized by $\Delta \chi^2_0$ the difference in $\chi^2$ between
the exact model fit and a constant flux fix to the light curve.
Each band is for a different planet/star radius
ratio $R_p/R_*$, and the width of the band shows the variation in
$\Delta\chi^2_{bc}/\Delta \chi^2_0$ for range of linear limb-darkening parameters
$u_1=0.0-0.4$.}
\label{fig:a1}

\end{figure}

\def\eq#1{equation (\ref{#1})}
\def\Eq#1{Eq.~\ref{#1}}
\def\kms{{\rm km\,s^{-1}}}
\def\AU{{\rm AU}}
\def\km{\rm km}
\def\au{\rm AU}
\def\kms{\rm km~s^{-1}}
\def\deg{^\circ}
\def\ntr{N_{\rm tr}}
\def\ttr{t_{\rm tr}}
\def\teq{t_{\rm eq}}
\def\tn{t_{\rm night}}
\def\pt{{\cal P}_{t}}
\def\pw{{\cal P}_{W}}
\def\psn{{\cal P}_{S/N}}
\def\ceq{\chi^2_{\rm eq}}
\def\cmin{\chi^2_{\rm min}}
\def\nn{N_{\rm n}}
\def\fl{f_{0}(\lambda)}
\def\fz{f_{0\lambda}}
\def\ptot{{\cal P}_{\rm tot}}

\appendix\label{chisqapx}

In this appendix, we derive the boxcar fitting algorithm used to
search for planetary transits.  The original developers of this
algorithm \citep{KOV02}, hereafter BLS, study its properties in the
presence of Gaussian noise.  We rederive the BLS method here in terms
of the familiar language of $\chi^2$ minimization, demonstrating the
equivalence of the two approaches (see also \citealt{ai04}).  In the
process, we elucidate the meaning of some of the detection statistics
introduced in \citet{KOV02}.  Finally, we quantify the accuracy of the
boxcar approximation to transit light curves and discuss appropriate
detection thresholds in the presence of correlated systematic errors
as well as pure noise.

We begin by considering a series of flux measurements, $F_k$, taken at
times $t_k$ with errors $\sigma_{F,k}$.  The equation,
${\overline{F}}\equiv (\sum_k F_k/\sigma^2_{F,k})/ (\sum_k
1/\sigma^2_{F,k})$, defines the mean error-weighted flux of the
measurements.  We can write the fractional deviations from the mean as
$m_k \equiv F_k/\, {\overline{F}}-1$ and the fractional errors as
$\sigma_k = \sigma_{F,k}/\, {\overline{F}}$.  For small deviations,
the difference from the mean in magnitudes approximates $-1.086m_k$.

A light curve of a star with transiting planet can be written as
$F(t)=F_0[1-\delta(t)]$, where $F_0$ is the unobscured flux of the
star, and $\delta(t)$ is the fractional drop in flux due to the
transiting planet as a function of time.  This generally depends on
five parameters, namely the planet period $P$, the phase of the planet
orbit $\phi$, the time it takes for the planet to cross one stellar
radius, the ratio of the planet radius to stellar radius $\rho$, and
the limb-darkening parameter of the star.  We can write the fractional
flux difference as $m(t)=F(t)/\, {\overline{F}}-1 = m_0 -
\delta(t)-m_0\delta(t)$, where we have defined $m_0=F_0/\,
{\overline{F}}-1$.  We will assume that $\delta\ll 1$ and $m_0\ll 1$,
and therefore $m(t)\simeq m_0-\delta(t)$.

For $\rho\ll 1$, and no limb darkening, the light curve $m(t)$ phased
to the period $P$ reduces to a simple boxcar, $\delta=\rho^2$ when the
planet is occulting the star and $\delta=0$ otherwise, with five
parameters: the unocculted flux $m_0$, the transit period $P$, phase
$\phi$, duration $\Delta t$, and depth $\delta$.  This simple form
allows an analytical solution for two of these parameters. For finite
$\rho$ and limb darkening, the expression for $\delta(t)$ increases in
complexity, but, as we show later, a simple boxcar still approximates
the variability.

The likelihood ${\cal L}$ that the model $m(t)$ with parameters
$\alpha=(m_0,\delta,P,\phi,\Delta t)$ describes the data $m_k$ is
simply,
\begin{equation}
-2\ln{\cal L} \equiv \chi^2 =  \sum_k \left[\frac{ m_k-m(t_k)}{\sigma_k}\right]^2.
\label{eqn:loglike}
\end{equation}
A likelihood maximization, or $\chi^2$ minimization, determines the
best-fit parameters.  Phasing the measurements $m_k$ to a given
period, we can split $\chi^2$ into two terms, including the points in
transit and points out of transit,
\begin{equation}
\chi^2 = \sum_i \left(\frac{ m_i-m_0 + \delta }{\sigma_i}\right)^2 +
\sum_j \left(\frac{ m_j-m_0}{\sigma_j}\right)^2,
\label{eqn:chi2}
\end{equation}
where the first sum over index $i$ is over the points in transit, and
the second over index $j$ is over the points out of transit.  The
quantity ${\cal D}$ of BLS corresponds to our $\chi^2/{\sum_k \sigma_k^{-2}}$, where the sum over index $k$ is over all points.  Since $\sum_k \sigma_k^{-2}$ is a constant, this
verifies the equivalence of their algorithm to a simple
$\chi^2$-minimization.  Expanding the quadratic terms in \eq{eqn:chi2}
yields,
\begin{equation}
\chi^2 = 
\sum_k \frac{m_k^2}{\sigma_k^2}
-2m_0 \sum_k \frac{m_k}{\sigma_k^2}
+m_0^2\sum_k \frac{1}{\sigma_k^2}
+2\delta\sum_i \frac{m_i}{\sigma_i^2}
-2\delta m_0  \sum_i\frac{1}{\sigma_i^2}
+\delta^2 \sum_i\frac{1}{\sigma_i^2}.
\label{eqn:chi2long}
\end{equation}
Given the identity
${\overline{m}}\equiv (\sum_k m_k/\sigma^2_{k})/ (\sum_k
1/\sigma^2_{k}) = 0$, it is clear that the first term is simply
the $\chi^2$ of a constant flux fit to the data, which we denote $\chi^2_0$,
and the second term is zero.  Therefore, the last four terms are the
improvement in $\chi^2$ between the constant-flux fit, and a boxcar
transit fit, for a given $P,\phi,\Delta t$.

Minimizing the expression for $\chi^2$ with respect to $m_0$, we find
$m_0=(\sum_i\sigma_i^{-2}/\sum_k \sigma_k^{-2})\delta$.
Inserting this into the expression for $\chi^2$, and then
minimizing with respect to $\delta$, we find
the parameters $\delta$,
$m_0$ that simultaneously minimize $\chi^2$ for a given $P,\phi,\Delta t$,
\begin{equation}
\delta = \frac{ -\sum_i m_i/\sigma_i^2}
{\sum_i \sigma_i^{-2}(1-{\sum_i \sigma_i^{-2}}/{\sum_k \sigma_k^{-2}})}
\end{equation}
\begin{equation}
m_0 = \frac{ -\sum_i m_i/\sigma_i^2}
{\sum_k \sigma_k^{-2}(1-{\sum_i \sigma_i^{-2}}/{\sum_k \sigma_k^{-2}})}.
\end{equation}

We note that the solution for $\delta$ does not impose a particular
sign.  The best-fit boxcar model may be of a transit or anti-transit
nature.  The improvement in $\chi^2$ at these best-fit parameters is

\begin{equation}
\Delta\chi^2 = \chi^2-\chi^2_0= \frac{ -\left(\sum_i m_i/\sigma_i^2 \right)^2}
{\sum_i \sigma_i^{-2}(1-{\sum_i \sigma_i^{-2}}/{\sum_k \sigma_k^{-2}})}.
\end{equation}
At fixed $P,\phi,\Delta t$, and for pure Gaussian noise, $\Delta \chi^2$ 
is distributed as $\chi^2$ with one degree of freedom (corresponding
to the one additional free parameter).   

The above expression gives the $\chi^2$ improvement for a particular
set of $P,\phi,\Delta t$.  The global solution that maximizes the
$\chi^{2}$ improvement, $\Delta \chi^{2}_{\rm max}$, requires a grid
search over the entire $P, \phi, \Delta t$ parameter space feasible for
transit detection.  In the BLS algorithm, evaluation of $\Delta\chi^2$
for a given $P,\phi,\Delta t$ amounts to simple error-weighted
binning, making the process extremely efficient and fast.  Due to the
large number of light curves with injected transits that must be
searched in order to determine the detection probability accurately
(see \S\ref{effcalc}), this study requires the excellent numerical
efficiency provided by the BLS method.  Comparison with BLS reveals
that $\Delta \chi^2_{\rm max}$ has a close correspondence with their
detection statistic SR, such that $\Delta \chi^2_{\rm max}/{\sum_k
\sigma_k^{-2}}=-({\rm SR})^2$.  If we assume constant noise, i.e.\
$\sigma_k={\rm constant}=\sigma$, and define $N_t$ to be the number of
points in transit, then the effective signal-to-noise ratio is ${\rm
S/N}=(|\Delta \chi^2_{\rm max}|)^{1/2}=N_t^{1/2}(\delta/\sigma)$.

Although, at fixed $P,\phi,\Delta t$, analytic expressions can provide
the significance of a given value of $\chi^2$ in the presence of pure
Gaussian noise, in the present case folding data over many trial
periods and phases hampers determination of the significance of the
global $\Delta \chi^2_{\rm max}$.  Thus, assessing the probability of
achieving the observed outcome by chance involves taking into account
the effective number of independent trials.  Several papers (e.g.,
BLS, \citealt{jcb02}) propose methods to address these issues.
However, as we have shown, in the presence of correlated systematic
errors, formulations with a Gaussian-noise basis underestimate the
detection thresholds needed to avoid falsely triggering on the
systematic errors.  Generally, one must use the properties of the data
themselves to set the appropriate detection threshold.

In general, a matched filter optimally detects a signal of known form
in noisy data.  However, the efficiency of boxcar fitting makes it
highly advantageous, and as we show next nearly optimal for transit
detection.  This is because, for small planets and modest
limb darkening, simple boxcars match well the planetary transit
signal.  We demonstrate this in Figure \ref{fig:a1}, where we show
$\Delta\chi^2_{bc}\equiv \chi^2_{bc}-\chi^2_{exact}$, the difference in 
$\chi^2$ between a boxcar fit to a planetary transit across a limb-darkened
star and the exact model fit, assuming uniform errors.   In
order to make the result independent of the error properties of the light curve,
we plot this normalized to $\Delta \chi^2_0 \equiv \chi^2_{constant}-\chi^2_{exact}$,
the difference in 
$\chi^2$ between  a constant flux fit to the light curve and the exact model fit .  
In other words, since $(\Delta \chi^2_0)^{1/2}$ is the total signal-to-noise
ratio of the transit detection, the ratio $\Delta\chi^2_{bc}/\Delta \chi^2_0$ is (the square
of) the fractional degradation of the signal-to-noise ratio that arises from fitting the approximate
boxcar form to the true transit shape.  We show $\Delta\chi^2_{bc}/\Delta \chi^2_0$ versus
the impact parameter of the transit for four different values of
$\rho=0.01,0.05,0.1$ and $0.2$, as well as linear limb-darkening
coefficients in the range $u_1=0-0.4$.  The fractional difference in
$\Delta\chi^2_0$ is less than $\sim 5\%$ for impact parameters $\la 0.8$ and
$\rho\la 0.1$, and it is always $\le 10\%$ for the range of parameters
relevant here.  However, if the number of detections is a strong
function of $\chi^2$ near the detection threshold, then even a $\sim
5\%$ change in $\chi^2$ can have a significant effect on the inferred
detection probability.  Therefore, it is important to inject realistic
transits when determining the recovery rate and detection
probability, to account for the imperfect match of the boxcar filter.

\begin{deluxetable}{rccccl}
\tablecaption{MDM 2.4m Observations\label{obsdat24}}
\tablehead{\colhead{Date (2001)} & \colhead{\#Exps} & \colhead{FWHM (arcsec)} & \colhead{Comments}}
\startdata
Oct. 24 & 75 & 1.2 & clear - 1st quarter moon\\
     25 & 73 & 1.4 & partly cloudy \\
     26 & 67 & 1.4 & partly cloudy \\
     27 & 22 & 1.6 & overcast  \\
     28 & 96 & 1.4 & cirrus \\
     29 & 86 & 1.3 & cirrus \\
     31 & 32 & 1.4 & partly cloudy \\
Nov.  1 & 32 & 1.4 & clear humid - full moon\\
      2 & 80 & 1.5 & clear - moon closest approach\\
      6 & 44 & 1.4 & clear humid \\
      7 & 92 & 1.3 & partly cloudy - 3rd quarter moon\\
      8 & 57 & 1.4 & cirrus    \\
     10 & 81 & 1.8 & cirrus    \\
     11 & 99 & 1.3 & clear   \\
\enddata
\end{deluxetable}

\clearpage
\begin{landscape} 
\begin{deluxetable}{ccccccccccccccccc}
\tabletypesize{\scriptsize}
\tablecaption{Transit candidate data\label{trncandprop}}
\tablehead{\colhead{ID} & \colhead{RA(2000.0)} & \colhead{Dec(2000.0)} & \colhead{$V$(mag)} & \colhead{\bv(mag)} & \colhead{$V-I$\, (mag)} & \colhead{$\chi^{2}_{\rm mem}$} & \colhead{$P$(d)} & \colhead{$\Delta f$(mag)} & \colhead{$\tau$(h)} & \colhead{$\phi$} & \colhead{$\Delta \chi^{2}/\Delta \chi^{2}_{\rm -}$} & \colhead{$\Delta \chi^{2}$} & \colhead{$f$} & \colhead{$M$(M$_{\odot}$)} & \colhead{Log((R/R$_{\odot}$)} & \colhead{Teff(K)}}
\startdata
30207 & 3:15:40.0 & +47:21:18 & 18.1 & 1.17 & 1.31 & 0.137 & 4.614 & 0.030 & 1.66 & 0.72 & 5.65 & 584   & 0.57 & \nodata & \nodata & \nodata \\
20513 & 3:15:04.6 & +47:15:09 & 18.6 & 1.10 & 1.27 & 0.028 & 1.637 & 0.018 & 4.71 & 0.95 & 3.85 & 840   & 0.49 & 0.91 & -0.095 & 5400 \\
20065 & 3:15:03.8 & +47:14:33 & 16.1 & 1.02 & 1.25 & 0.418 & 3.026 & 0.115 & 4.18 & 0.24 & 3.82 & 88086 & 0.59 & \nodata & \nodata & \nodata \\
20398 & 3:14:49.5 & +47:16:03 & 18.4 & 1.29 & 2.00 & 0.863 & 0.349 & 0.145 & 0.90 & 0.72 & 3.76 & 66056 & 0.29 & \nodata & \nodata & \nodata \\
20274 & 3:14:35.9 & +47:19:29 & 19.3 & 1.67 & 3.37 & 4.390 & 0.302 & 0.050 & 0.98 & 0.48 & 3.23 & 14063 & 0.21 & \nodata & \nodata & \nodata \\
70718 & 3:13:56.2 & +47:06:59 & 21.1 & 1.28 & 1.79 & 0.017 & 0.640 & 0.032 & 3.07 & 0.16 & 2.86 & 312   & 0.22 & 0.63 & -0.253  & 4300 \\    
\enddata
\end{deluxetable}
\clearpage
\end{landscape}
\end{document}